\documentstyle[12pt,aaspp4]{article}

\newcommand\kms{\ifmmode{~\rm km\th s^{-1}}\else ~km\th s$^{-1}$\fi}
\newcommand\th{\thinspace}
\def\ubv{\hbox{$U\!BV$}}
\def\by{\hbox{$b\!-\!y$}}

\lefthead{Torres et al.}
\righthead{Absolute dimensions of GG Ori}

\begin{document}

\title{Absolute dimensions of the unevolved B-type eclipsing binary
GG~Orionis\altaffilmark{1}}

\altaffiltext{1}{Some of the observations reported here were obtained
with the Multiple Mirror Telescope, a joint facility of the
Smithsonian Institution and the University of Arizona.}

\author{Guillermo Torres}
\affil{Harvard-Smithsonian Center for Astrophysics, 60 Garden St.,
Cambridge, MA 02138}
\authoremail{gtorres@cfa.harvard.edu}

\author{Claud H. Sandberg Lacy\altaffilmark{2}}
\altaffiltext{2}{Visiting astronomer, Kitt Peak National Observatory,
National Optical Astronomy Observatories, operated by the Association
of Universities for Research in Astronomy, Inc., under cooperative
agreement with the National Science Foundation.}
\affil{Department of Physics, University of Arkansas, Fayetteville,
AR 72701}
\authoremail{clacy@comp.uark.edu}

\author{Antonio Claret}
\affil{Instituto de Astrof\'\i sica de Andaluc\'\i a, CSIC, Apartado
3004, E-18080 Granada, Spain}
\authoremail{claret@iaa.es}

\author{Jeffrey A. Sabby\altaffilmark{2}}
\affil{Department of Physics, University of Arkansas, Fayetteville,
AR 72701}
\authoremail{}

\vskip 2in\centerline{To appear in \emph{The Astronomical Journal}, December 2000}

\newpage

\begin{abstract}

We present photometric observations in $B$ and $V$ as well as
spectroscopic observations of the detached, eccentric 6.6-day
double-lined eclipsing binary GG~Ori, a member of the Orion OB1
association. Absolute dimensions of the components, which are
virtually identical, are determined to high accuracy (better than 1\%
in the masses and better than 2\% in the radii) for the purpose of
testing various aspects of theoretical modeling. We obtain $M_A =
2.342 \pm 0.016$~M$_{\sun}$ and $R_A = 1.852 \pm 0.025$~R$_{\sun}$ for
the primary, and $M_B = 2.338 \pm 0.017$~M$_{\sun}$ and $R_B = 1.830
\pm 0.025$~R$_{\sun}$ for the secondary. The effective temperature of
both stars is $9950 \pm 200$~K, corresponding to a spectral type of
B9.5. GG~Ori is very close to the ZAMS, and comparison with current
stellar evolution models gives ages of 65-82~Myr or 7.7~Myr depending
on whether the system is considered to be burning hydrogen on the main
sequence or still in the final stages of pre-main sequence
contraction. Good agreement is found in both scenarios for a
composition close to solar.  We have detected apsidal motion in the
binary at a rate of $\dot\omega = 0.00061 \pm
0.00025$~deg~cycle$^{-1}$, corresponding to an apsidal period of $U =
10700 \pm 4500$~yr. A substantial fraction of this ($\sim$70\%) is due
to the contribution from General Relativity, and our measurement is
entirely consistent with theory. The eccentric orbit of GG~Ori is well
explained by tidal evolution models, but both theory and our
measurements of the rotational velocity of the components are as yet
inconclusive as to whether the stars are synchronized with the orbital
motion. 
	
\end{abstract}

\keywords{binaries: eclipsing --- binaries: spectroscopic --- stars:
evolution --- stars: fundamental parameters --- stars: individual
(GG~Orionis)}

\newpage

\section{Introduction}

The discovery of GG~Orionis (HD~290842, Tycho~4767~857~1, $V =
10.4-11.1$, \ion{B9.5}{5}, $\alpha = 05^h 43^m 10\fs2$, $\delta =
-00\arcdeg 41\arcmin 15\arcsec$, epoch and equinox J2000) as a
variable star is due to \markcite{h34}Hoffmeister (1934), who observed
the object photographically at the Sonneberg Observatory. The
correct period of 6.631 days was first given by
\markcite{k51}Kordylewski (1951) based on visual and photographic
times of minimum. This author obtained a mean visual light curve, and
established that the orbit is eccentric from the displacement of the
secondary minimum. 

Aside from the occasional measurement of the times of eclipse by a
number of authors, GG~Ori has remained until recently a rather
neglected system.  Double lines in the spectrum were detected by
\markcite{l84}Lacy (1984), who described them as being narrow and of
nearly equal strength, but no detailed spectroscopic study has been
made to date. The first photoelectric light curves were published by
\markcite{z97}Zakirov (1997), who presented light elements for this
well detached binary solved by the method of \markcite{l93}Lavrov
(1993). 

Based on the fact that the orbit is eccentric, it is expected that the
system may present a measurable apsidal motion.  This effect is of
great interest in the study of detached eclipsing binaries because it
provides information on the internal structure of stars that may be
compared with predictions from theory.  From its spectral type and
other known properties, GG~Ori was listed by \markcite{g85}Gim\'enez
(1985) as a good candidate for the study of the contribution of
General Relativity to the secular displacement of the line of apsides,
given that the relativistic effect is expected to be dominant in this
particular case. 

GG~Ori is located in the Orion~OB1 association (see, e.g.,
\markcite{b64}Blaauw 1964; \markcite{wh77}Warren \& Hesser 1977), a
complex region of star formation that has been the subject of numerous
studies to determine the properties of the population of young stars
and surrounding gas. The binary is located not far from the Belt of
Orion, and therefore there is reason to expect that the system might
also be quite young, adding to its interest. 

In this paper we present new high-quality photoelectric light curves
in two passbands, which we analyze together with other published
photometry. We also report the results of our intensive spectroscopic
monitoring of GG~Ori that, combined with the light curves, enable us
to derive highly precise absolute dimensions for both components of
the system. The stars turn out to be nearly identical in all their
properties. 

Our current knowledge of the internal structure and evolution of stars
is such that observed stellar properties determined with errors of
even 5\% are of little value to constrain the models in any
significant way, since the differences between competing calculations
or the effects of slight changes in the input physics or atomic
constants are below this level (see Andersen \markcite{a91}1991,
\markcite{a98}1998).  The masses we obtain here have errors smaller
than 1\%, and the radii are determined to better than 2\%.  We use
these determinations for a comparison with the predictions of recent
stellar evolution models, which suggest a very young age for the
system.  An analysis of all available eclipse timings along with our
radial velocities leads to a small but apparently significant apsidal
motion detection. We discuss the importance of the relativistic effect
mentioned above compared to classical effects from rotational and
tidal distortions.  In addition, we examine the predictions from tidal
theory regarding spin-orbit synchronization and the circularization of
the orbit. 

\section{Spectroscopic observations and reductions}

GG~Ori was observed spectroscopically at the Kitt Peak National
Observatory (KPNO) and at the Harvard-Smithsonian Center for
Astrophysics (CfA). Observations at KPNO were made with the 2.1m and
coud\'e-feed spectrometer from 1983 December to 1999 March. A total of
15 spectrograms were obtained with a variety of CCD detectors.  A
spectral resolution of 0.02~nm or 2 pixels was used before May 1998,
and 0.03~nm or 3 pixels after that date. The spectral coverage was
about 10~nm in the first 12 spectrograms, and 32~nm in the last 3
spectrograms. The central wavelength of those observations was 450~nm. 

Projected rotational velocities ($v \sin i$) were determined by
comparing line widths of the binary components' unblended features
with artificially-broadened features in the spectra of comparison
stars with known values of $v \sin i$. These comparison stars were
chosen to be of nearly the same spectral type as GG~Ori, and were
observed with the same instrumental configuration as the binary. The
reference stars used were $o$~Peg (HR~8641, \ion{A1}{4}, $v \sin i =
10\kms$; Fekel 1998, private communication), 68~Tau (HR~1389,
A2 IV-V, $v \sin i = 18\kms$; \markcite{h82}Hoffleit 1982), and
HR~8404 (21~Peg, \ion{B9.5}{5}, $v \sin i = 4\kms$;
\markcite{f99}Fekel 1999).  From these comparisons in 10 of the
spectrograms, values of $24 \pm 2\kms$~and $23 \pm 2\kms$~were
determined for the primary and secondary in GG~Ori, respectively,
based mainly on the \ion{Mg}{2} 448.1~nm line.  The uncertainties
given account for the scatter of the line width measurements as well
as the agreement between the results using different standard stars.
The component of GG~Ori we call here the ``primary" (also star ``A")
is the one eclipsed at phase 0.0 in the light curve (see \S4).
Formally it is also the more massive one, but only marginally so, as
we describe later. 

Radial velocities of the components were measured by cross-correlation
with the {\tt FXCOR} task in IRAF\footnote{IRAF is distributed by the
National Optical Astronomy Observatories, which is operated by the
Association of Universities for Research in Astronomy, Inc., under
contract with the National Science Foundation.}, using standard stars
(same as above) with known radial velocities as templates.  Radial
velocities for the low-amplitude spectroscopic binary star $o$~Peg
were taken from the orbit of \markcite{f99}Fekel (1999).  The radial
velocity of HR~8404 was taken as $+0.2\kms$~based on 12 velocities by
\markcite{f99}Fekel (1999), and for 68~Tau we adopted the value
$+39.0\kms$, from the same source. Our measurements for GG~Ori from
the 11 spectra with relatively unblended lines are given in Table~1. 

In addition to the radial and rotational velocities, we determined the
line strength ratio from 17 line pairs in spectra of good
signal-to-noise (S/N) ratio. This may be used as a proxy for the light
ratio between the stars. We obtained ${\rm EW}_A/{\rm EW}_B = 1.01 \pm
0.02$, corresponding to the blue region of the spectrum. Because the
two components are virtually identical, no correction for the
difference in temperature is needed, and the ratio of the equivalent
widths, ${\rm EW}_A/{\rm EW}_B$, can be assumed to be identical to the
light ratio. 

Our observations of GG~Ori at the CfA were collected mostly with the
echelle spectrograph on the 1.5m Tillinghast reflector at the F.\ L.\
Whipple Observatory on Mt.\ Hopkins (Arizona), where the system was
monitored from 1996 March to 2000 March. Occasionally we observed with
an identical spectrograph on the Multiple Mirror Telescope (also atop
Mt.\ Hopkins, Arizona).  A total of 42 spectra were recorded with a
photon-counting Reticon detector covering a single echelle order
centered at 518.7~nm. The resolution is 0.015~nm
($\lambda/\Delta\lambda \sim 35,\!000$), and the spectra span about
4.5~nm. The S/N ratios range from about 25 to 40 per resolution
element. The zero point of the velocity system was monitored by means
of nightly exposures of the dusk and dawn sky, in the manner described
by \markcite{l92}Latham (1992). The accuracy of the CfA velocity
system, which is within about 0.1\kms~of the reference frame defined
by minor planets in the solar system, is documented in the previous
citation and also by \markcite{slt99}Stefanik et al.\ (1999). 

Radial velocities were determined using the CfA implementation of
TODCOR (\markcite{zm94}Zucker \& Mazeh 1994), a two-dimensional
cross-correlation algorithm that uses two templates, one for each
component of the binary. The templates were selected from a large
library of synthetic spectra based on the latest model atmospheres by
R.\ L.\ Kurucz, computed for us specifically for the wavelength of our
observations (Morse \& Kurucz 1998, private communication). The
instrumental profile is explicitly included by applying a Gaussian
convolution with a FWHM of 8.5\kms, corresponding to our spectral
resolution.  These synthetic spectra have been calculated over a wide
range of effective temperatures, projected rotational velocities,
metallicities, and surface gravities. Initially we used templates with
a temperature of $T_{\rm eff} = 10000$~K for both stars, surface
gravity of $\log g = 4.5$ (cgs), and solar metallicity.  The projected
rotational velocities of the components were determined by running
extensive grids of correlations against templates covering a range of
values of $v \sin i$, seeking the best match to the observed spectra
as determined by the correlation coefficient averaged over all
exposures.  We obtained $v \sin i = 16 \pm 1$\kms~for both components,
which is significantly different than our determination based on KPNO
spectra. We discuss this further in \S5 and \S6. For the final
templates we adopted a temperature of 9950~K for both stars and $\log
g = 4.3$, based on the results from the light curve solutions and
spectroscopic orbits presented below. 

Systematic errors in the radial velocities resulting from line
blending are always a concern, particularly when the goal is to
achieve the highest possible precision and accuracy in the mass
determinations.  Although the use of a two-dimensional correlation
technique such as TODCOR in principle minimizes those errors (see
\markcite{l96}Latham et al.\ 1996), experience has shown us that this
must be checked on a case-by-case basis, particularly in view of the
narrow spectral window of the CfA observations (e.g., Torres et al.\
\markcite{t97}1997; \markcite{t00}2000). For this we generated
artificial composite spectra by adding together synthetic spectra for
the two components with Doppler shifts appropriate for each actual
time of observation, computed from a preliminary orbital solution. We
adopted also a light ratio close to that for the real stars (see
below).  We then processed these simulated spectra with TODCOR in the
same manner as the real spectra, and compared the input and output
velocities. The differences derived in this way vary systematically
with phase (or radial velocity), as expected, and were applied to the
real velocities as corrections, even though in the case of GG~Ori they
are quite small ($\leq 0.5$\kms) and hardly affect the results. The
final radial velocities from the CfA spectra with the corrections
included are given in Table~2. 
	
The light ratio was derived from these observations using TODCOR as
described by \markcite{zm94}Zucker \& Mazeh (1994). Because of the
relatively small number of lines in the 4.5~nm spectral window
observed at CfA, significant errors can be introduced in the light
ratio when features of each component shift in and out of this region
by up to 0.2~nm due to orbital motion. To estimate and correct for
this effect we followed a procedure analogous to that described above
for the velocities, and found the magnitude of the systematic error to
be about 3.5\%.  After correcting for this we obtain $L_A/L_B = 1.05
\pm 0.03$ at a mean wavelength of 518.7~nm, which is sufficiently
close to the visual band that we will assume it is the light ratio in
$V$, given that the stars are virtually identical. 
	
Preliminary double-lined orbital solutions based on our radial
velocities were computed separately for the KPNO and CfA measurements,
and are compared in Table~3. Initially we adopted the eclipse
ephemeris given in the General Catalog of Variable Stars (GCVS)
(\markcite{k85}Kholopov 1985): Min~I (HJD) $ = 2,\!433,\!596.496 +
6.63147 \cdot E$. However, the residuals from the orbit for the CfA
data showed a systematic pattern and the scatter was significantly
larger than expected from previous experience with similar
spectroscopic material. This suggested that the ephemeris is not
accurate enough to be propagated forward by $\sim50$~yr (2600 cycles)
to the epoch of our velocity measurements. Therefore, a new linear
ephemeris computed from times of minimum collected from the literature
was derived (see next section) and used here.  The possibility of a
systematic difference between the center-of-mass velocities of the
primary and secondary components was explored by solving for an offset
simultaneously with the orbital elements. This was done independently
for the two data sets. No shift was expected because the stars are so
similar. The results in the sense $\langle$primary minus
secondary$\rangle$ were $+1.42 \pm 0.97\kms$~for KPNO and $-0.74 \pm
0.53\kms$~for CfA. Since these are indeed not significantly different
from zero, the velocities for the two components can be considered for
all practical purposes to be on the same reference frame. 
	
As seen in Table~3, there are slight differences in some of the
elements of these two orbital solutions, in particular in the velocity
amplitude of the secondary component ($K_B$), but we do not consider
them to be significant in view of the small number of KPNO
observations (only 11). Therefore, for the final solution discussed
below we have merged the two data sets. 

\section{Apsidal motion analysis and final spectroscopic solution}

As indicated earlier, the noticeable eccentricity of the orbit of
GG~Ori suggests the possibility of apsidal motion, even though none
has been reported for this system. In extreme examples this can affect
the spectroscopic solutions quite significantly, as in the case of
V477~Cyg (\markcite{p68}Popper 1968).  Before computing the final
orbital fit, we therefore searched the literature for eclipse timings,
both to improve the ephemeris and to investigate the apsidal motion. 

Table~4 lists all times of minimum available to us, of which there are
25 primary minima and 32 secondary minima covering about 65 yr
(1930-1995). The majority were determined by visual or photographic
means, and only the 6 recent photoelectric timings and one visual
timing have published uncertainties. For the rest we determined the
errors iteratively, based on the mean residuals for each type of
observation from preliminary fits: $\sigma_{\rm pg} = 0.038$ days for
the photographic minima, and $\sigma_{\rm v} = 0.028$ days for the
visual minima. One visual estimate was found to give a large residual
(0.14 days, $\sim5\sigma$), and was excluded. It is indicated in
parentheses in Table~4. 

A linear ephemeris fit to these data gives Min~I (HJD) $ =
2,\!449,\!717.66253(21) + 6.6314948(16) \cdot E$, with a phase
difference between the primary and secondary minima of $\Delta\Phi =
0.42252 \pm 0.00024$. This is the ephemeris we used above in our
preliminary orbital solutions, and the period is very similar to that
reported by \markcite{z97}Zakirov (1997) based on a smaller number of
observations. Although no unusual trends were seen in the timing
residuals, we investigated the possibility of apsidal motion by
computing the apparent periods separately for the primary and the
secondary. We obtained $P_I = 6.6314971 \pm 0.0000021$ days and
$P_{II} = 6.6314890 \pm 0.0000030$ days, which are different at the
2.2$\sigma$ level. Next we submitted these same data to an apsidal
motion analysis using the method by \markcite{lc92b}Lacy (1992b). For
this we assumed a fixed eccentricity equal to the average of the
values derived in columns (2) and (3) of Table~3, and a fixed
inclination angle of $89\arcdeg$ from preliminary light curve
solutions. The result is only marginally significant ($\dot\omega =
0.00050 \pm 0.00027$ deg~cycle$^{-1}$), largely due to the weak
constraint provided by the early photographic and visual times of
minimum, which have relatively low weight. Changes in those weights
have little effect. 

Our radial velocities extend the time base provided by the times of
minimum by another 5 yr, and contain valuable information on the
longitude of periastron that can also be used. In addition, they
provide a much stronger constraint on the eccentricity, which must
otherwise be fixed when using only eclipse timings. The optimal
solution, therefore, is to combine both kinds of measurements into a
single least squares fit, solving simultaneously for the spectroscopic
orbital elements and the apsidal motion. In this way the information
on the period contained in the times of minimum is implicitly taken
into account for the spectroscopic elements, rather than having to
adopt a fixed ephemeris as we have done in \S2. At the same time, the
effect of the possible rotation of the line of apsides on the
spectroscopic elements is also accounted for. In addition, we have
allowed for an arbitrary shift between the KPNO and CfA velocity
systems in view of the difference in the way the two zero points were
established. The offset turns out to be negligible:
$\langle$CfA$-$KPNO$\rangle$~$= -0.06 \pm 0.39\kms$. 

The results of this simultaneous fit are given in column (4) of
Table~3. The relative weights of the primary and secondary velocities
in each data set, as well as the weights of the times of minimum, have
been iterated until reaching convergence for a reduced $\chi^2$ of
unity. Tests show once again that the weights assigned to the visual
and photographic timings, or to the velocities, have little effect on
the results.  The velocity residuals from this solution are listed
separately in Table~1 and Table~2, and the timing residuals are given
in Table~4.  The final spectroscopic orbital solution is represented
graphically in Fig.~1, along with the observations. 

The apsidal motion resulting from the final fit is $\dot\omega =
0.00061 \pm 0.00025$ deg~cycle$^{-1}$, which is significant at the
$2.5\sigma$ level, and differs from our preliminary estimate above by
less than half of its uncertainty.  Also listed in Table~4 are the
sidereal and anomalistic periods ($P_s$ and $P_a$), as well as the
apsidal motion period $U = 10700 \pm 4500$ yr. Periastron passage
occurs at a photometric phase of 0.06. A plot of the $O\!-\!C$
deviations of the times of minimum from the linear terms of the
apsidal motion is given in Fig.~2 along with the predicted deviations.
The top panel draws attention to the fact that the observations cover
less than 1\% of a full cycle, while the bottom panel expands the
region around the observations, with the coverage provided by the
radial velocity measurements being indicated in the upper right. 

\section{Photometric observations and light curve solutions}

Differential light curves were obtained by CHSL at the Cerro Tololo
Inter-American Observatory (CTIO) with the 24-inch Lowell telescope in
the Johnson $B$ and $V$ bands during 1993-1995. Absolute indices were
also measured separately based on measurements that were carefully
tied to the standard $\ubv$ system through observations of secondary
standards from \markcite{l73}Landolt (1973), typically 30-40 standards
per night.  The absolute photometry was made at Mount Laguna
Observatory near San Diego (California) in the fall season of 1989,
and at CTIO in 1988-1990 and 1993-1995. The procedures that were used
are described in detail by \markcite{lc92a}Lacy (1992a). The $\ubv$
indices on the standard system are shown in Table~5, and the 257
differential $BV$ observations are given in Table~6 and Table~7. The
differential magnitudes are in the sense
$\langle$variable$-$comparison$\rangle$ and are referred to
BD~$-1\arcdeg$1013 (HD~38165, \ion{B9}{5}).  The comparison stars (see
Table~5) were found to be constant at a level of about 0.008~mag for
the standard deviation of the differences between comparison stars.
The precision of the CTIO differential observations, based on previous
results, is estimated to be about 0.006~mag in both $B$ and $V$ at the
magnitudes of the program stars. 

In addition to the absolute photometric indices mentioned above, which
we have computed separately for the two observing intervals, $uvby$
indices are available from \markcite{hh75}Hilditch \& Hill (1975). The
results from the three different sources are somewhat inconsistent.
The $V$ magnitude is slightly brighter by about 2.5\% in the 1988-1990
results compared to the 1993-1995 results and the value of
\markcite{hh75}Hilditch \& Hill (1975), which is $V = 10.380 \pm
0.015$. The $\bv$ color index during 1993-1995 is considerably bluer
than the value measured in 1988-1990.  These discrepancies cannot be
explained by differences in sky conditions (which were always
photometric), nor by differences in aperture size, centering errors,
or seasonal variations.  Furthermore, the scatter from the light curve
fits described later in this section is significantly larger than
expected based on our previous results with the equipment used at CTIO
at these magnitude levels. The standard deviation of the residuals
from the fitted photometric orbit of FS~Mon (\markcite{lc00}Lacy et
al.\ 2000), for example, was 0.006 and 0.005~mag in $B$ and $V$,
respectively, whereas the corresponding figures for GG~Ori (see below
for details) are 0.014 and 0.009~mag.  Both binaries were observed
during the same runs.  Intrinsic variability of one or both of the
components of GG~Ori is therefore suspected. 

Differential light curves in the Johnson $U$, $B$, $V$, and $R$ bands
have also been obtained in Uzbekistan by the group led by M.\ Zakirov.
They used 0.6m telescopes at the Maidanak Observatory, in the southern
part of Uzbekistan. Their differential magnitudes in the sense
$\langle$variable$-$comparison$\rangle$ relative to the star
BD~$-1\arcdeg$1013 have been described by \markcite{z97}Zakirov
(1997). The transformation of these measurements to the standard
system is somewhat less secure than for our own observations, but
there are marginal indications that the $\bv$ index of GG~Ori became
bluer by a few hundredths of a magnitude in the interval 1993-1994,
toward the end of their observations. Though far from being conclusive
evidence of intrinsic variability, we note that this is in the same
direction as the trend suggested in Table~5.  We have re-analyzed the
differential observations by Zakirov with the same methods we use for
our own measurements. 

The light curve fitting was done with the NDE model as implemented in
the code EBOP (\markcite{e81}Etzel 1981; \markcite{pe81}Popper \&
Etzel 1981), and the ephemeris adopted is that of \S3.  The main
adjustable parameters are the ratio of the central surface brightness
of the secondary star ($J_B$) in units of that of the primary, the
relative radius of the primary ($r_A$) in units of the separation, the
ratio of the radii ($k \equiv r_B/r_A$), the inclination of the orbit
($i$), and the geometric factors $e\cos\omega$ and $e\sin\omega$ which
account for the orbital eccentricity. As usual, we have allowed also
for a photometric scale factor and a phase shift in all solutions.
Auxiliary quantities needed in the analysis include the
gravity-brightening coefficients, for which we adopt the values 1.00,
0.85, 0.70, and 0.59 in $U$, $B$, $V$, and $R$, respectively, from
\markcite{m73}Martynov (1973). For the reflection coefficients we
adopted the value 1.0, as appropriate for stars with radiative
envelopes.  Limb-darkening coefficients ($u$) were included as one of
the variables to be fitted, although in one case (noted below) it was
fixed to an appropriate value from the tables of \markcite{wr85}Wade
and Rucinski (1985). The mass ratio ($q\equiv M_B/M_A = 0.9982$) was
adopted from the spectroscopic analysis in \S3. 

Preliminary solutions revealed a few outliers with residuals greater
than $5\sigma$ (one of the CTIO measurements in $B$, two in $V$, and
one of Zakirov's measurements in each of his four bands), which were
given zero weight in subsequent iterations. 

The fitting procedure converged in the general solutions of the CTIO
light curves with all variables adjusted simultaneously (see columns 2
and 3 in Table~8).  As is often the case in the analysis of partially
eclipsing binary stars with nearly equal components, the fitted values
of the parameter $k$, the ratio of radii, are rather uncertain, but
near unity.  Attempts at general solutions of the Zakirov light curves
failed to converge.  In cases such as this light ratios estimated from
spectrograms are very important for constraining the photometric
solutions.  For GG~Ori we have at our disposal two spectroscopic
determinations of the light ratio, one approximately in the $B$ band
(from KPNO) and the other close to the $V$ band (from CfA). A grid of
EBOP solutions was run on the CTIO light curves for fixed values of
$k$ between 0.94 and 1.10, in steps of 0.01. Within this range of
about 15\%, the rms residuals from the fits were found to change very
little as a function of $k$ ($\leq 0.00005$~mag in $B$ and $\leq
0.00006$~mag in $V$), supporting our concerns about the indeterminacy
of $k$ from the light curves alone.  The sum of the relative radii
changes by less than 0.7\% over the entire range of $k$ values.  The
light ratios in $B$ and $V$ were computed from each of these
solutions, and then by interpolation we determined the $k$ values that
correspond to our measured spectroscopic light ratios: $k = 0.996 \pm
0.011$ from the KPNO spectra, and $k = 0.972 \pm 0.015$ from the CfA
spectra (see Fig.~3). The value we adopt is the weighted average,
$\langle k\rangle = 0.988 \pm 0.009$.  For comparison, the average
ratio of the radii derived from the CTIO light curve solutions without
the constraint from spectroscopy is $k = 1.03 \pm 0.04$ (see Table~8),
which is different by $1\sigma$. 

With the ratio of the star sizes fixed to the weighted average given
above, we repeated the light curve fits separately for the CTIO and
\markcite{z97}Zakirov (1997) photometry, and the adopted solutions are
given in Table~8 and Table~9. Attempts to include third light as a
parameter showed that, within uncertainties of about 1\%, it was not
significantly different from zero. 

The CTIO observations in $B$ and $V$ and the best fit models are shown
in Fig.~4, and the data by \markcite{z97}Zakirov (1997) in $\ubvr$
along with the corresponding fitted light curves are shown in Fig.~5.
Table~10 gives the weighted averages of the light elements in each
band obtained from the two data sets. In addition to $r_A$ and $k$, we
list also the sum of the relative radii, $r = r_A + r_B$, and the
radius of the secondary, $r_B$. The errors for these quantities were
derived as described in the Appendix. There is good agreement between
the photometric determinations of $e \cos\omega$ and $e \sin\omega$
and the corresponding spectroscopic values, which are $-0.1200 \pm
0.0017$ and $+0.1865 \pm 0.0020$, respectively. 

The results in Table~10 show that the components of GG~Ori are nearly
indistinguishable in size and of virtually the same luminosity, which
is consistent with the fact that they are also very similar in mass
(\S3). The departure from the spherical shape is insignificant.  Both
eclipses are partial, with approximately 91\% of the light of the
primary blocked at phase 0.0 (which is a transit), and 90\% of the
light of the secondary eclipsed at the other minimum. The
earlier study by \markcite{z97}Zakirov (1997) reported the primary
eclipse to be total (larger secondary star blocking the primary),
although the ratio of the radii was much closer to 1.0. 

The uncertainties assigned to the adjusted quantities in Table~10 are
generally the internal errors produced by EBOP. As is well known, the
formal uncertainties in least-squares solutions in which one or more
quantities have been held fixed are typically too small because
correlations between the elements are artificially eliminated (as we
have done by fixing $k$, for example). The quantities $k$, $r_A$, and
$i$ are the most important for determining the absolute dimensions of
the components.  The error in $k$ is based on the uncertainty in the
spectroscopic light ratios, and therefore does not suffer from this
shortcoming. For $r_A$ and $i$ we have increased the
formal uncertainties to account for all contributions to the error. 

The rms deviations given in Table~8 for the CTIO observations are
somewhat larger than expected, based on previous experience, and
suggest that one or both components of GG~Ori may be intrinsically
variable. We have searched for patterns in the residuals by computing
the power spectra in the $B$ and $V$ bands, but no significant
periodicities were found. 
	
\section{Absolute dimensions}

The combination of the spectroscopic results in Table~3 and the light
curve results in Table~10 leads to the absolute masses and radii for
GG~Ori, shown in Table~11. The masses are determined with a precision
of 0.7\%, and the radii are good to about 1.4\%. 

The fact that the surface brightness ratio $J_B$ is near unity over a
wide range of wavelengths indicates that the effective temperatures of
the stars are essentially the same. From $J_B$ in the $V$ band the
difference in visual surface brightness is $\Delta F^{\prime}_V =
0.0002 \pm 0.0004$ (\markcite{p80}Popper 1980), which translates into
a completely negligible color difference $\Delta(\bv) = 0.0005 \pm
0.0009$, and a formal temperature difference of only 8~K. 
	
The discrepancies in the absolute photometry at different epochs
pointed out earlier in \S4 complicate the derivation of the effective
temperature of the components. Because the light variations may be
real, we have chosen to adopted the straight average of the largest
and smallest $\ubv$ indices for GG~Ori in Table~5: $V = 10.372 \pm
0.013$, $\bv = 0.511 \pm 0.021$, $\ub = 0.324 \pm 0.004$. The errors
for the first two of these quantities are simply half of the
difference between the 1988-1990 and 1993-1995 measurements.  The
reddening was derived using the reddening-free index $Q = (\ub) - 0.72
(\bv)$ (\markcite{jm53}Johnson \& Morgan 1953) and the calibration by
\markcite{dds76}Deutschman, Davis \& Schild (1976). We obtained
$(\bv)_0 = -0.036 \pm 0.006$, $(\ub)_0 = -0.070 \pm 0.020$, $E(\bv) =
0.547 \pm 0.022$, and $A_V = 1.696 \pm 0.068$ (using $A_V = 3.1\cdot
E(\bv)$).  The formal errors do not account for uncertainties in the
calibrations themselves, which are difficult to quantify.  These color
indices correspond to the combined light, but since the components are
virtually identical they are also the indices of the individual stars. 

An independent measure of the reddening may be obtained from the
$uvby$ photometry by \markcite{hh75}Hilditch \& Hill (1975): $\by =
0.408 \pm 0.015$, $c_1 = 1.002 \pm 0.030$, $m_1 = +0.016 \pm 0.020$.
The calibration by \markcite{c78}Crawford (1978) for B-type stars
leads to $(\by)_0 = -0.030$ and $E(\by) = 0.438$, from which $E(\bv) =
0.592$. This is slightly larger than the reddening derived from the
$\ubv$ photometry. However, GG~Ori is near the end of the range of
validity of the $uvby$ calibrations, where they become rather
uncertain.  Also, we note that the $(\by)_0$ index corresponds to a
spectral type of B8.5 according to Table~1 by \markcite{p80}Popper
(1980). But the fact that we see no sign of the \ion{He}{1} 447.1~nm
line in our KPNO spectra of the object strongly suggests that the
spectral type cannot be earlier than B9.5.  We have therefore chosen
to rely only on the Johnson photometry above. 

The de-reddened $\bv$ index corresponds to a temperature of $9950 \pm
200$~K, and a spectral type of B9.5 (\markcite{p80}Popper 1980).
Further properties of the stars are listed in Table~11, including the
mean density ($\bar\rho$), the absolute visual magnitudes, and the
distance. The latter two are based on the visual surface brightness
parameter $F^{\prime}_V = 3.974 \pm 0.003$ derived from the intrinsic
color $(\bv)_0$ and the tabulation by \markcite{p80}Popper (1980), and
are thus independent of bolometric corrections. 

The distance modulus of GG~Ori, $m\!-\!M = 8.20 \pm 0.10$ ($d = 438
\pm 20$~pc), is similar to other determinations for various subregions
of the Orion~OB1 complex (\markcite{wh77}Warren \& Hesser 1977;
\markcite{bgz94}Brown, de Geus, \& de Zeeuw 1994;
\markcite{bwb99}Brown, Walter \& Blaauw 1999; \markcite{z99}de Zeeuw
et al.\ 1999), and lends support to its association. 

Also listed in Table~11 are the projected rotational velocities
expected if the axial rotations were synchronized with the mean
orbital motion of the binary and with the orbital motion at periastron
(pseudo-synchronization). The two independent measurements of $v \sin
i$ in \S2 based on our KPNO and CfA spectra disagree, with the KPNO
results being closer to the pseudo-synchronous values, while the CfA
determinations suggest synchronization with the mean orbital motion.
We discuss this further in \S6 in the context of tidal evolution
theory. 

No metallicity determination is available for GG~Ori. A rough estimate
can be obtained from our CfA spectra following a procedure analogous
to that used to derive $v \sin i$, and the result is [m/H]~$= -0.15
\pm 0.20$, consistent with the solar abundance. 
		
\section{Discussion}

In this section we compare the properties of GG~Ori as listed in
Table~11 with predictions from models. The three aspects of theory we
focus on are stellar evolution, internal structure, and tidal
evolution. For each of these we require calculations of the basic
properties of the components, which we have taken from evolutionary
tracks computed specifically for the exact masses we determined for
the stars. The evolutionary code we used is that by
\markcite{c95}Claret (1995).  Further details on the input physics are
described by \markcite{cg92}Claret \& Gim\'enez (1992). Convection in
these models is treated with the standard mixing-length prescription,
with a fixed mixing-length parameter of 1.52$H_p$ that gives the best
fit between a solar model and the observed properties of the Sun. A
moderate amount of core overshooting is assumed ($\alpha_{\rm ov} =
0.20 H_p$), although in the case of GG~Ori this effect is
insignificant due to the unevolved nature of the system. 

\subsection{Stellar evolution}

Consistent with the young environment surrounding the object (the
Orion OB1 association), we find that GG~Ori is indeed very close to
the Zero Age Main Sequence (ZAMS). This is illustrated in Fig.~6,
which shows a greatly enlarged section of the evolutionary tracks as
they approach the main sequence.  The full extent of the main sequence
band is shown in the inset.  The binary is so close to the ZAMS that,
given the uncertainties, it is difficult to determine whether the
stars are already burning hydrogen in their cores, or whether they are
still in the final stage of contraction towards the main sequence.
The error bars shown reflect the uncertainties in the effective
temperatures of the components and their surface gravities.  The error
in the placement of the tracks due to the uncertainty in the measured
masses is indicated in the lower left corner. 

The models plotted in Fig.~6 correspond to the case where the system
is already on the main sequence (MS). Tracks for two different
chemical compositions ($Z$, $Y$) are shown, giving metallicities very
close to solar ([m/H] $= +0.03$, $dY/dZ = 1.65$, solid lines; and
[m/H] $= +0.05$, $dY/dZ = 1.95$, dashed lines) consistent with the
indication in \S5. The enrichment laws ($dY/dZ$) implied by the best
fit values of the helium abundance ($Y$) are within the range found in
other determinations (see, e.g., \markcite{p93}Peimbert 1993;
\markcite{r94}Renzini 1994; \markcite{pp98}Pagel \& Portinari 1998;
\markcite{it98}Izotov \& Thuan 1998), and are also similar to the
values favored by other eclipsing binaries (\markcite{r00}Ribas et
al.\ 2000), using the same set of evolutionary models. 

Our attempts to find evolutionary tracks that fit the location of
GG~Ori assuming it is nearing the end of the pre-main sequence (PMS)
phase also gave acceptable results (see Fig.~7), considering the error
bars, although the agreement is formally not as good as the
main-sequence case. The PMS tracks shown in the figure are for [m/H]
$= +0.07$ and $dY/dZ = 2.09$. 

The evolutionary ages derived for the components are quite different
in the two scenarios (see Table~12). In the main sequence case the
average age for the two stars is 82~Myr (for $Z = 0.020$, $Y =
0.2729$) or 65~Myr (for $Z = 0.021$, $Y = 0.2810$), and the difference
in age between the components is similar to the formal uncertainty,
which is derived from the error in $\log g$. On the other hand, if
GG~Ori is on the PMS the models give an average age of only 7.7~Myr.
Few cases are known of eclipsing binaries with well-determined
absolute dimensions in which at least one component is a bona-fide PMS
star: EK~Cep (\markcite{p87}Popper 1987; \markcite{cgm95}Claret,
Gim\'enez, \& Mart\'\i n 1995), TY~CrA (\markcite{c98}Casey et al.\
1998), and possibly RS~Cha (\markcite{cn80}Clausen \& N\"ordstrom
1980; \markcite{p97}Pols et al.\ 1997; \markcite{m00}Mamajek, Lawson,
\& Feigelson 2000).  Such systems are particularly valuable to test
models of PMS evolution, where theory remains essentially unchallenged
by observations so far. The stars in GG~Ori are so similar and so
close to the ZAMS, however, that the constraint on PMS evolution is
not very strong.  Unevolved systems such as this are much more useful
for testing opacity and metallicity effects in the models when coupled
with an accurate spectroscopic determination of the metal abundance,
which unfortunately is not yet available for GG~Ori.  Whether or not
the system has already settled on the main sequence, it is undoubtedly
young. In this connection we note, incidentally, that light variations
such as those hinted at in \S4 are not entirely unexpected. 

It is a remarkable coincidence that among the eclipsing systems with
the best known absolute dimensions, no fewer than \emph{six} have at
least one component with virtually the same mass as the stars in
GG~Ori, to within 1\%. These systems are V451~Oph, YZ~Cas,
$\beta$~Aur, WX~Cep, SZ~Cen (\markcite{a91}Andersen 1991), and
V364~Lac (\markcite{t99}Torres et al.\ 1999). Thus, a total of eight
stars (including GG~Ori) can in principle be compared with \emph{a
single evolutionary track}, and they span the entire main sequence
band from the ZAMS (GG~Ori) to the shell hydrogen burning phase
(SZ~Cen, age $\sim$800~Myr). In Fig.~8 we show this comparison against
one of the models used above for GG~Ori ($Z = 0.020$, $Y = 0.2729$),
for a composition close to solar.  The agreement with theory is very
good, indicating that all these systems are well represented by a
single metallicity. 
	
\subsection{Internal structure and General Relativity}

Our detection of the apsidal motion of the binary in \S3 provides the
opportunity to test models of the internal structure of stars.  In
addition, the relatively short orbital period and large masses of the
components along with the fact that the orbit is eccentric led
\markcite{g85}Gim\'enez (1985) to propose GG~Ori as a good candidate
for the study of the general-relativistic (GR) contribution to the
apsidal motion. At the current level of the uncertainties in the
measurement of apsidal motion in binaries, the theoretical GR
contribution is separable from the classical (Newtonian) terms due to
the gravitational quadrupole moment induced by rotation and tides. The
total apsidal motion can then be expressed as $\dot\omega_{tot} =
\dot\omega_N + \dot\omega_{GR}$. The two effects cannot be measured
separately, though, so we have chosen here to compare the total
theoretical value with the measured quantity. 

In the case of GG~Ori, the GR term is predicted to be about 2.5 times
larger than the classical term, contributing $\sim$70\% to
$\dot\omega_{\rm tot}$.  Following \markcite{g85}Gim\'enez (1985), we
obtain $\dot\omega_{GR} = 0.000454 \pm 0.000002$~deg~cycle$^{-1}$. 

To compute $\dot\omega_N$ we have used the internal structure
constants from theory (including the higher order terms $k_3$ and
$k_4$), which are calculated at each point along the evolutionary
track of each component.  The contribution of the rotational
distortions to the Newtonian apsidal motion rate $\dot\omega_N$
depends on the ratio between rotational angular velocity and the
Keplerian angular velocity. This ratio is often derived under the
assumption that the stars are already synchronized at periastron.  For
a system as young as GG~Ori this may not necessarily be the case, and
in fact the observational evidence we have is conflicting (\S5; see
also \S6.3).  We have therefore relied on our own measurements of the
projected rotational velocities, $v \sin i$, even though our two
independent determinations from KPNO and CfA spectra are somewhat
different (\S2).  The KPNO values ($v_A \sin i = 24 \pm 2\kms$, $v_B
\sin i = 23 \pm 2\kms$) lead to $\dot\omega_N = 0.000209 \pm
0.000012$~deg~cycle$^{-1}$, while the CfA measurements ($v \sin i = 16
\pm 1\kms$ for both components) give $\dot\omega_N = 0.000183 \pm
0.000009$~deg~cycle$^{-1}$. 

When adding this to the GR term, we obtain $\dot\omega_{tot} =
0.000663 \pm 0.000012$~deg~cycle$^{-1}$ (KPNO) and $\dot\omega_{tot} =
0.000637 \pm 0.000010$~deg~cycle$^{-1}$ (CfA), in which the errors
account for all contributions from measured quantities.  These
predictions are to be compared with the observed value,
$\dot\omega_{obs} = 0.00061 \pm 0.00025$~deg~cycle$^{-1}$ (\S3).
There is good agreement with theory, although the uncertainty in the
observed value is large enough that the constraint on the models is
very weak. Further measurements of times of minimum are required to
improve $\dot\omega_{obs}$. 

\subsection{Tidal evolution}

Tidal forces within a binary tend to synchronize the rotation of each
component with the orbital motion, and to make the orbit circular. The
two main mechanisms proposed to describe these effects
(\markcite{tt97}Tassoul \& Tassoul 1997, and references therein;
\markcite{z92}Zahn 1992, and references therein) make somewhat
different predictions for the timescales of these processes. 
In this section we compare our observations with both, although the
hydrodynamical mechanism by Tassoul has often been found to be too
efficient (see, e.g., \markcite{cgc95}Claret, Gim\'enez, \& Cunha
1995).  Synchronization times and circularization times have been
computed as described in the latter reference and also by
\markcite{cc97}Claret \& Cunha (1997), by integrating the differential
equations describing the evolution of the rotation and eccentricity.
All time-dependent properties of the stars were interpolated directly
from the evolutionary tracks for each component. 

In performing these calculations for young stars such as GG~Ori a
number of complications arise regarding the pre-main sequence phase
where the radii of the stars are much larger than on the main
sequence. One of them is that in principle the orbital period (or
semimajor axis) is also changing during this phase. For example, if
the period of GG~Ori is assumed to be fixed at the current value of
6.6~days, we find that the size of the orbit is too small to
accommodate the very large radii predicted for the stars during the
early stages of contraction along the Hayashi tracks. Thus the
equations for the orbital evolution (period and eccentricity) and
rotational evolution are coupled and must be integrated simultaneously
(see, e.g., \markcite{dmm92}Duquennoy, Mayor, \& Mermilliod 1992). 

\markcite{zb89}Zahn \& Bouchet (1989) have done this for late type
stars (0.5~$M_{\sun} \leq M \leq 1.25~M_{\sun}$) and concluded that
all binaries with periods up to 7 or 8 days arrive on the main
sequence with their orbits already circularized, and that further
evolution of the eccentricity on the main sequence is negligible.
However, eccentric systems on the main sequence with periods shorter
than this \emph{are} actually observed (see, e.g.,
\markcite{m92}Mathieu et al.\ 1992).  For earlier-type binaries such
as GG~Ori the efficiency of the tidal mechanisms may be different, and
a detailed investigation of the evolution in the PMS phase is beyond
the scope of this paper. 

Tidal theory is still largely in development, and many aspects of it
remain somewhat controversial.  Nevertheless, useful comparisons with
the observations are still possible under certain simplifying
assumptions, and have indeed been made in a number of cases.
\markcite{vp95}Verbunt \& Phinney (1995), \markcite{cgc95}Claret,
Gim\'enez, \& Cunha (1995), \markcite{cc97}Claret \& Cunha (1997), and
others, have compared the predictions of the main mechanisms mentioned
above with the properties of detached eclipsing systems with
well-determined absolute dimensions.  Their calculations were done in
detail (i.e., by \emph{integrating} the corresponding differential
equations rather than simply using the \emph{timescales} for
syncronization and circularization), but avoided the PMS problem by
excluding it altogether (\markcite{vp95}Verbunt \& Phinney 1995) or by
including only the final stages of contraction, and assumed also that
the orbital period does not change significantly, which may be a valid
approximation when the stars are already close to the ZAMS. 

We have followed a similar prescription here and considered only the
final loop in the evolutionary tracks preceding the ZAMS (see Fig.~8).
We start our integrations at the onset of the nuclear reaction
$^{12}{\rm C}(p,\gamma)~^{13}{\rm N}(\beta+\nu)~^{13}{\rm
C}(p,\gamma)~^{14}{\rm N}$.  At this stage this is not the only source
of energy, since contraction is still taking place.  The
synchronization and circularization times ($t_{sync}$, $t_{circ}$) are
not strongly dependent on the chemical composition in the range we
have considered above. Using the formalism by Tassoul we obtain $\log
t_{sync} = 6.875$ for both components of GG~Ori. The predictions from
the mechanism by Zahn, on the other hand, give $\log t_{sync}^A =
8.888$ and $\log t_{sync}^B = 8.890$, which are significantly longer.
The circularization times are nearly identical in both theories:
$t_{circ} = 8.888$ (Tassoul) and $t_{circ} = 8.889$ (Zahn). These
values are to be compared with the mean age of the system counted from
the starting point of the integrations, which differs by 5.4~Myr from
the true ages given in \S6.1. The modified age is then $\log t =
7.884$ ($Z = 0.020$, $Y = 0.2729$) or $\log t = 7.775$ ($Z = 0.021$,
$Y = 0.2810$), assuming GG~Ori is already burning hydrogen on the main
sequence. 

We illustrate this in a slightly different way in Fig.~9, where we
focus on $\log g$ as a sensitive measure of stellar evolution. The
value of the surface gravity at which synchronization or
circularization is achieved ($\log g_{\rm crit}$) is shown as a
function of orbital period for the two main tidal theories. Stars
evolve upwards in this diagram. The measured values of $\log g$ for
the primary and secondary of GG~Ori are represented by crosses, with
the error bars being of the same size as the symbols. According to the
theory by Zahn the stars should not yet have reached synchronous
rotation, which agrees with the measured $v \sin i$ values from our
CfA spectra. Also, the orbit should not yet be circular, as is indeed
observed.  The theory by Tassoul also predicts that the orbit should
still be eccentric, but indicates that the stars should already be
rotating synchronously with the orbital motion, which appears to be
more consistent with the $v \sin i$ values from our KPNO spectra.
Although as mentioned above the mechanism by Tassoul is often found to
be too efficient, a more accurate measurement of the projected
rotational velocities in GG~Ori could shed more light on this issue. 
	
\section{Conclusions}

New photometric and spectroscopic observations of the eccentric binary
GG~Ori combined with a reanalysis of data from the literature have
allowed us to derive definitive orbital parameters and physical
properties of the components. Our determinations have formal errors
smaller than 1\% in the masses and smaller than 2\% in the radii.
GG~Ori thus joins the elite of stars with the best established
absolute dimensions.  The system is very young, and in fact it is so
close to the ZAMS that within the observational errors the comparison
with current stellar evolution models is unable to tell us whether it
has already settled on the main sequence (age $\sim 65$--82~Myr,
depending on the metallicity), or whether it is still at the end of
the contraction phase (age $=7.7$~Myr). We find good agreement with
both scenarios for a chemical composition close to solar. 

The tentative detection of apsidal motion in GG~Ori with a period of
$U = 10700$~yr has allowed us to make an initial test of interior
structure models. This system is particularly interesting in that the
contribution from General Relativity should be substantial ($\sim$70\%
of the total apsidal motion). Although the agreement with theory is
good, further observations to improve the error in $\dot\omega$ are
necessary for a more stringent test. 

Current mechanisms that describe the tidal evolution of binary
properties disagree as to whether the rotation of the components of
GG~Ori should be synchronized with the orbital motion. Unfortunately
our own measurements of $v \sin i$ have not settled the issue.
Although this area of theory is perhaps the weakest of the comparisons
we have discussed, the models do explain the significant eccentricity
of the orbit system at the relatively short period of 6.6~days. 
	
\acknowledgments

We thank P.\ Berlind, M.\ Calkins, D.\ W.\ Latham, A.\ Milone, and R.\
P.\ Stefanik, who obtained many of the spectroscopic observations used
here, and R.\ Davis, who maintains the CfA database of radial
velocities. We are also grateful to Daryl Willmarth at KPNO for
assistance with the spectroscopic observations there.  Paul Etzel
provided helpful comments on aspects of the light curve solutions. We
thank the referee for useful suggestions.  This research has made use
of the SIMBAD database, operated at CDS, Strasbourg, France, and of
NASA's Astrophysics Data System Bibliographic Services. 

\section{Appendix}

The determination of the error in the relative radius of the secondary
component, $r_B$, in eclipsing binaries has often been a source of
confusion among users of EBOP. Results are occasionally published in
which that uncertainty is derived by propagating errors in the
expression $r_B = k\cdot r_A$, where $k$ (defined as $r_B/r_A$), the
primary radius $r_A$, and their associated uncertainties are usually
adopted directly from the output of the program.  Aside from the fact
that the formal errors in the latter two quantities are typically
underestimated (see \S4), this procedure is incorrect, as already
pointed out clearly in the documentation for EBOP (\markcite{e80}Etzel
1980). The reason is that $k$ and $r_A$ are, as a rule, strongly
(negatively) correlated, and simply adding their uncertainties by
quadratures does not account for this correlation. It is still
possible to make correct use of the expression above if the
correlation is explicitly taken into account by adding the
corresponding term from the off-diagonal elements of the covariance
matrix in the least squares solution.  However, this information is
not readily available in the standard output from EBOP.  Alternative
procedures for estimating $\sigma_{r_B}$ are discussed in the
documentation for the code (see also \markcite{pe81}Popper \& Etzel
1981). For example, when $k$ is fixed in the least squares solution
one may use as an approximation the expression
 \begin{equation}
 \sigma_{r_B}= k\cdot \sigma_{r_A},
 \end{equation}
 although this explicitly ignores the error in $k$, which in our case
is well known independently of EBOP. 

The reason for the strong correlation between $k$ and $r_A$ is that
the \emph{sum} of the radii, $r \equiv r_A + r_B$, which is directly
related to the form of the light curve, is usually very well
determined, and therefore if $k$ is increased the primary radius $r_A$
must decrease to maintain the sum. This suggests that a more sensible
way to approach the problem might be to consider the expressions
 \begin{equation}
 r_A = \left({1\over 1+k}\right)r~,~~~~~~~~~r_B = \left({k\over 1+k}\right)r~,
 \end{equation}
 in which the independent variables are $r$ and $k$, which are
\emph{not} strongly correlated in general. For GG~Ori we have shown
this to be the case in \S4. Propagation of errors in these equations
may therefore be expected to give more reasonable estimates of
$\sigma_{r_A}$ and $\sigma_{r_B}$:
 \begin{equation}
 \sigma_{r_A} = {1\over 1+k}\left[\left({r\over 1+k}\right)^2 \sigma_k^2 + \sigma_r^2\right]^{1/2}~,
 ~~~~\sigma_{r_B} = {1\over 1+k}\left[\left({r\over 1+k}\right)^2 \sigma_k^2 + k^2 \sigma_r^2\right]^{1/2}.
 \end{equation}
 Note that $\sigma_{r_B}$ will be smaller or larger than
$\sigma_{r_A}$ depending only on the value of $k$, as in eq.(1). In
fact, setting $\sigma_k = 0$ and combining the expressions above leads
back to (1). 

Practical application of these equations requires an estimate of
$\sigma_r$.  Although $r$ is not one of the adjustable variables in
EBOP, it is not difficult to obtain an estimate of its error by
experimenting with the light curve solutions and assessing the
sensitivity of $r$ to the various input quantities, as we have in fact
done with $r_A$. However, to ensure consistency with the error in
$r_A$, we have chosen to use the first of the equations in (3) to
solve for $\sigma_r$, since $\sigma_{r_A}$ is already known and so is
$\sigma_k$. Once we determined the value of $\sigma_r$ needed to
reproduce $\sigma_{r_A}$, we used it in the expression for
$\sigma_{r_B}$.

\clearpage

\begin{figure}
\plotfiddle{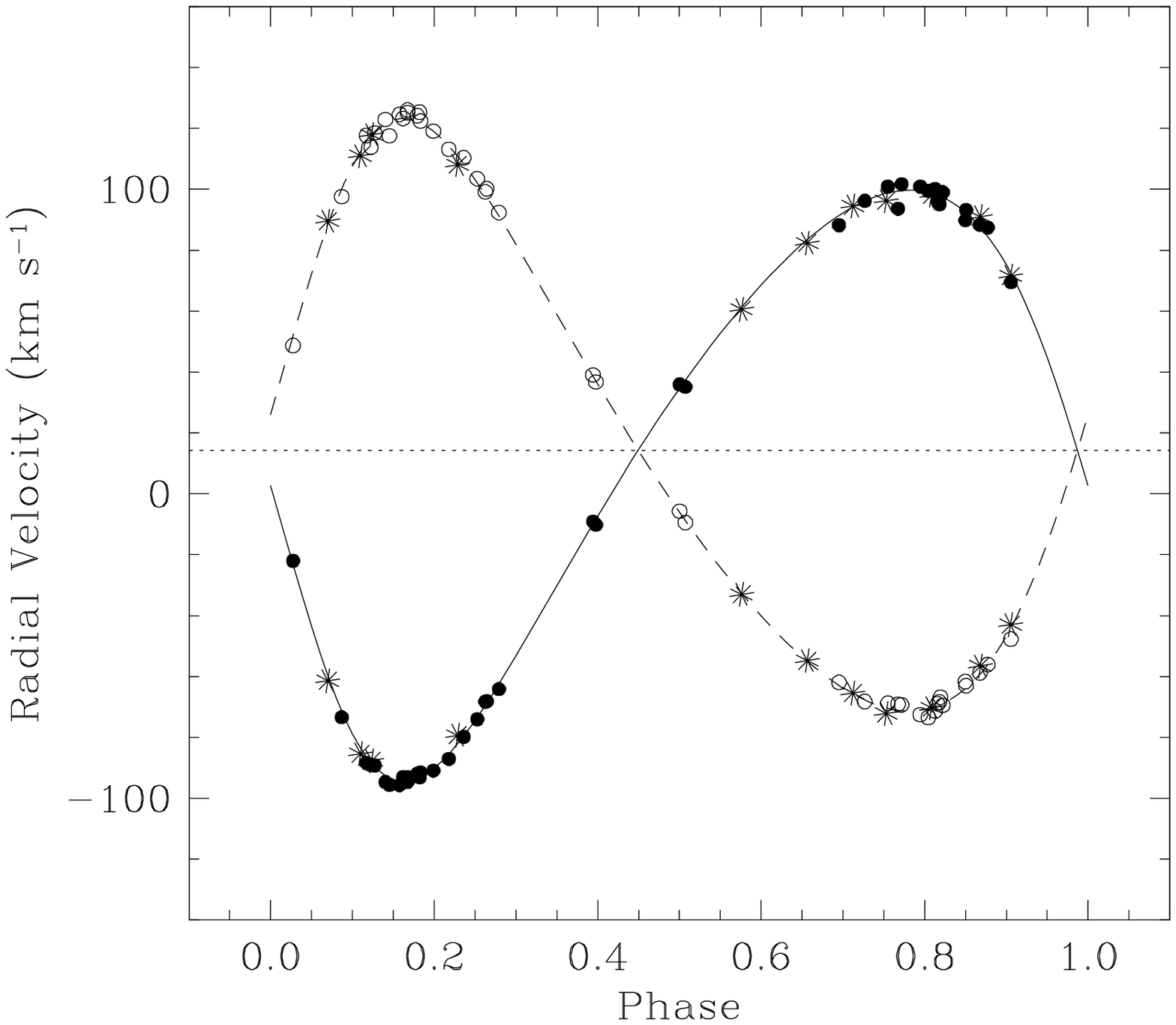}{4in}{0}{80}{80}{-240}{-150}
\caption{Radial-velocity curves of GG~Ori from our
final spectroscopic solution that includes times of minimum in the
least-squares fit. The filled circles are for the CfA observations of
the primary, the open circles are for the secondary, and the asterisks
represent the velocity measurements from KPNO.\label{fig1}}
\end{figure}

\clearpage

\begin{figure}
\plotfiddle{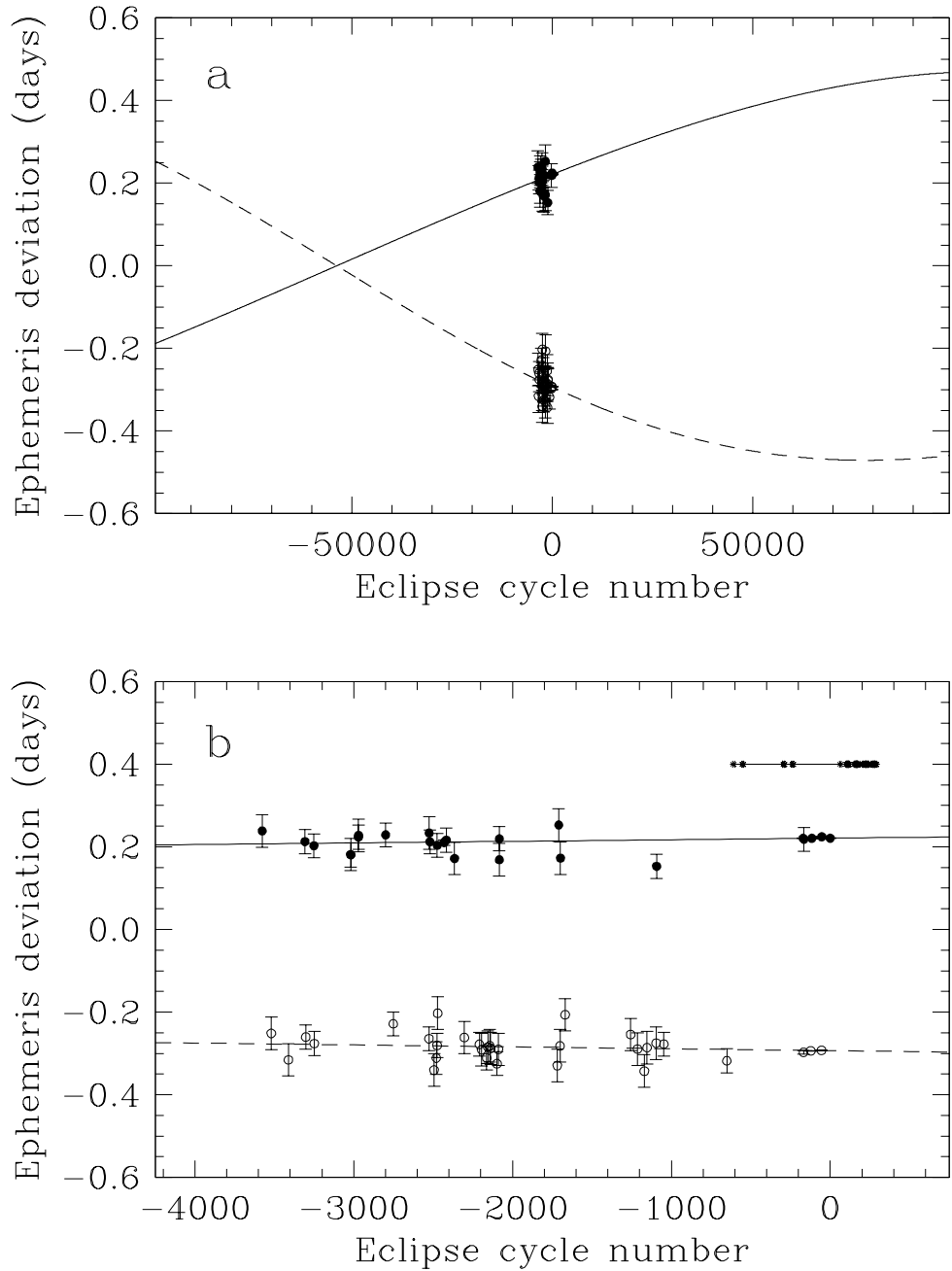}{7in}{0}{80}{80}{-245}{-75}
\caption{a) Ephemeris curve for GG~Ori along with
all available times of eclipse (solid line for the primary); b)
enlargement to show the distribution of the measurements. The time
distribution of the spectroscopic observations included in the apsidal
motion analysis is indicated by the horizontal line in the upper
right.}
\end{figure}

\clearpage

\begin{figure}
\plotfiddle{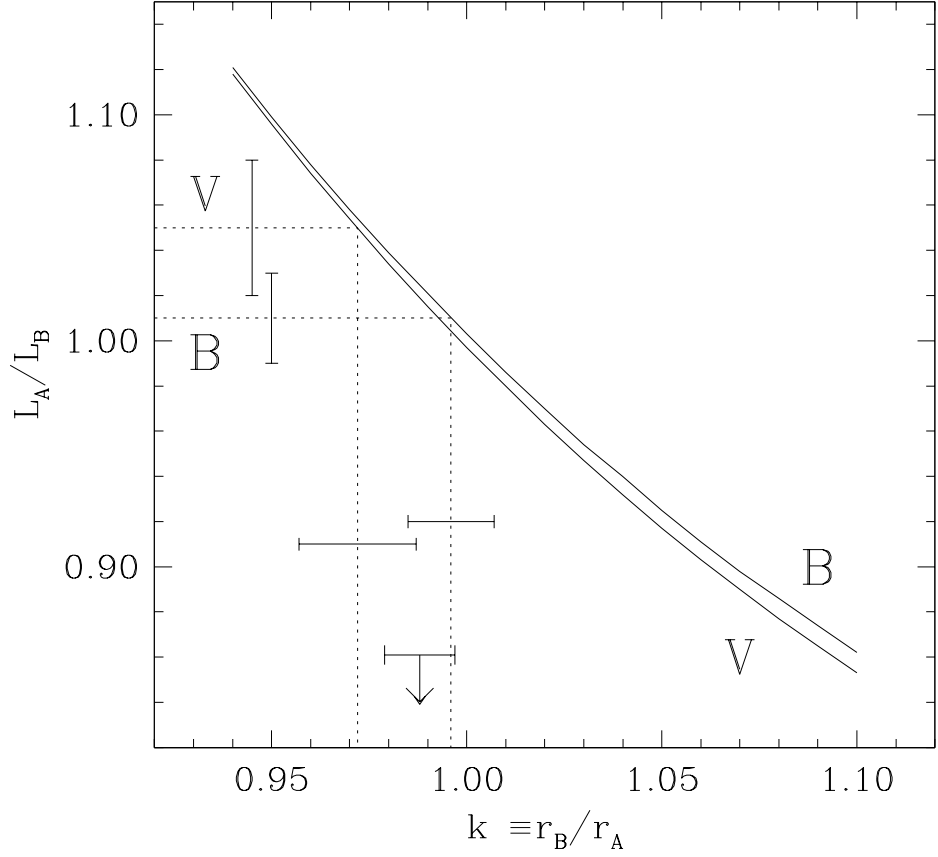}{4in}{0}{80}{80}{-270}{-150}
\caption{Application of the spectroscopic
constraints on the light ratio in $B$ and $V$ to determine $k$, the
ratio of the radii of the components. The solid curves are derived
from light curve solutions based on our CTIO photometry with $k$ held
fixed. The arrow with the error bar marks the resulting weighted
average of $k$.}
\end{figure}

\clearpage

\begin{figure}
\plotfiddle{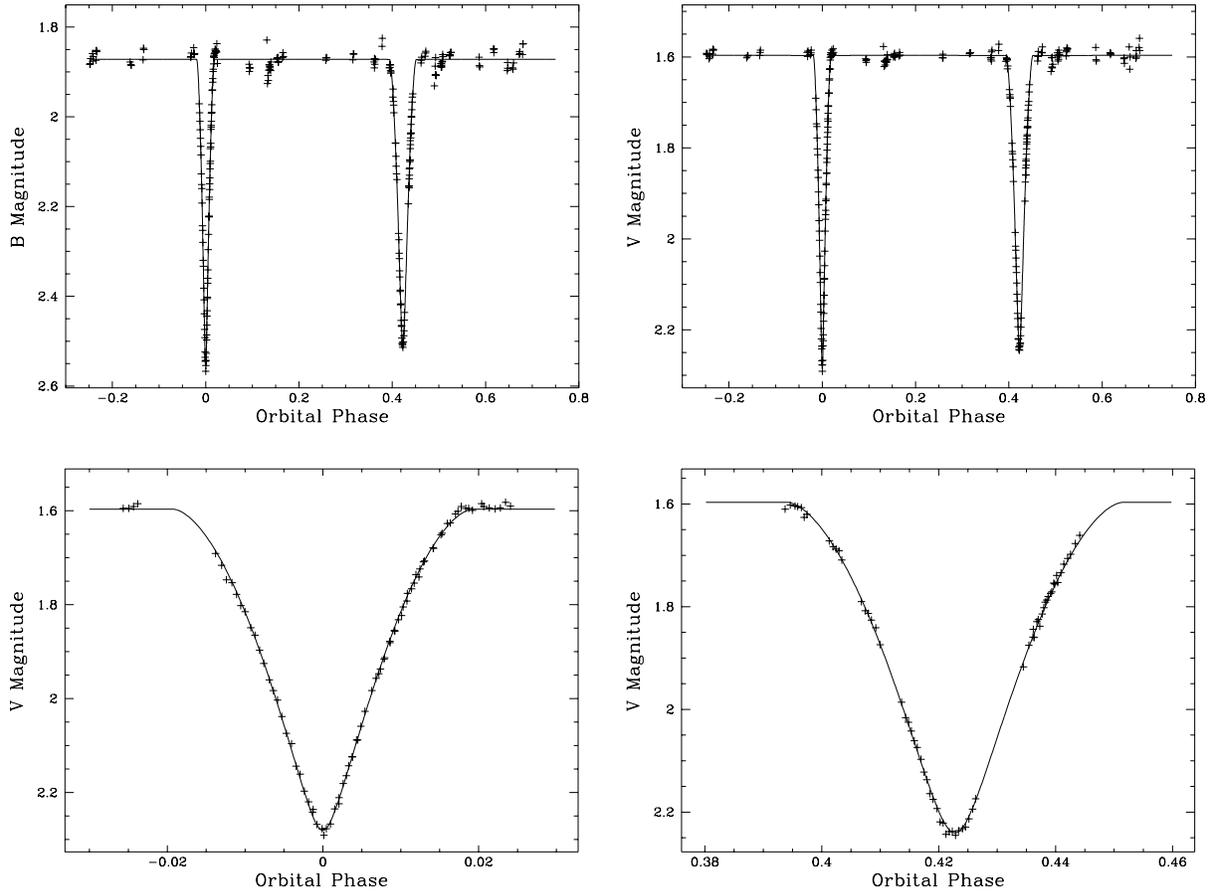}{4in}{0}{100}{100}{-300}{-220}
\caption{Light curves obtained at CTIO in the $B$
and $V$ bands along with the fitted solutions from EBOP. The lower
panels expand the region near the primary and secondary minima for the
$V$ band.}
\end{figure}

\clearpage

\begin{figure}
\plotfiddle{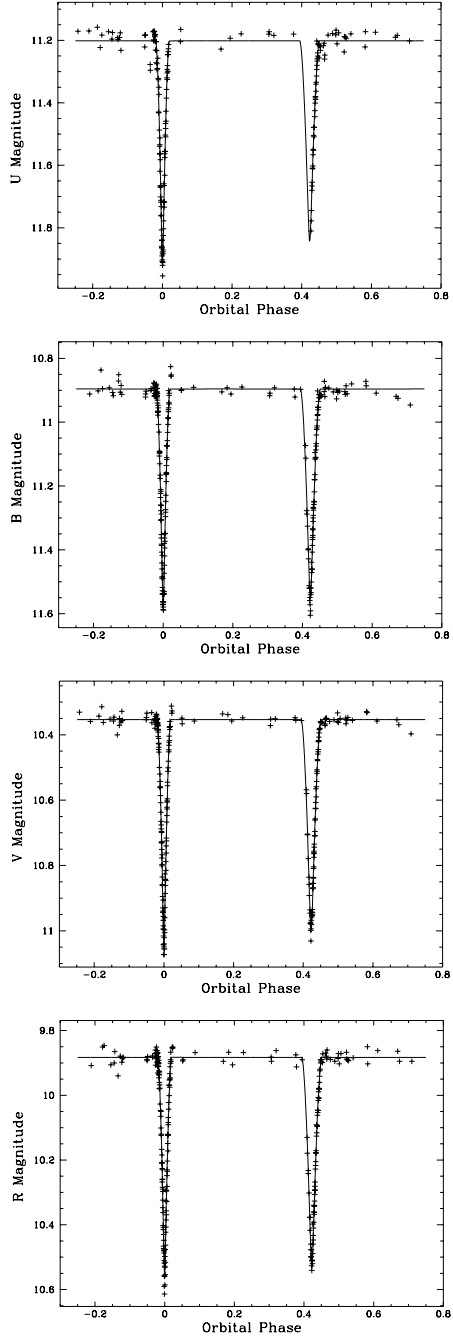}{7in}{0}{80}{80}{-225}{-70}
\caption{Light curves in $\ubvr$ based on the
observations by Zakirov (1997), along with our fits.}
\end{figure}

\clearpage

\begin{figure}
\plotfiddle{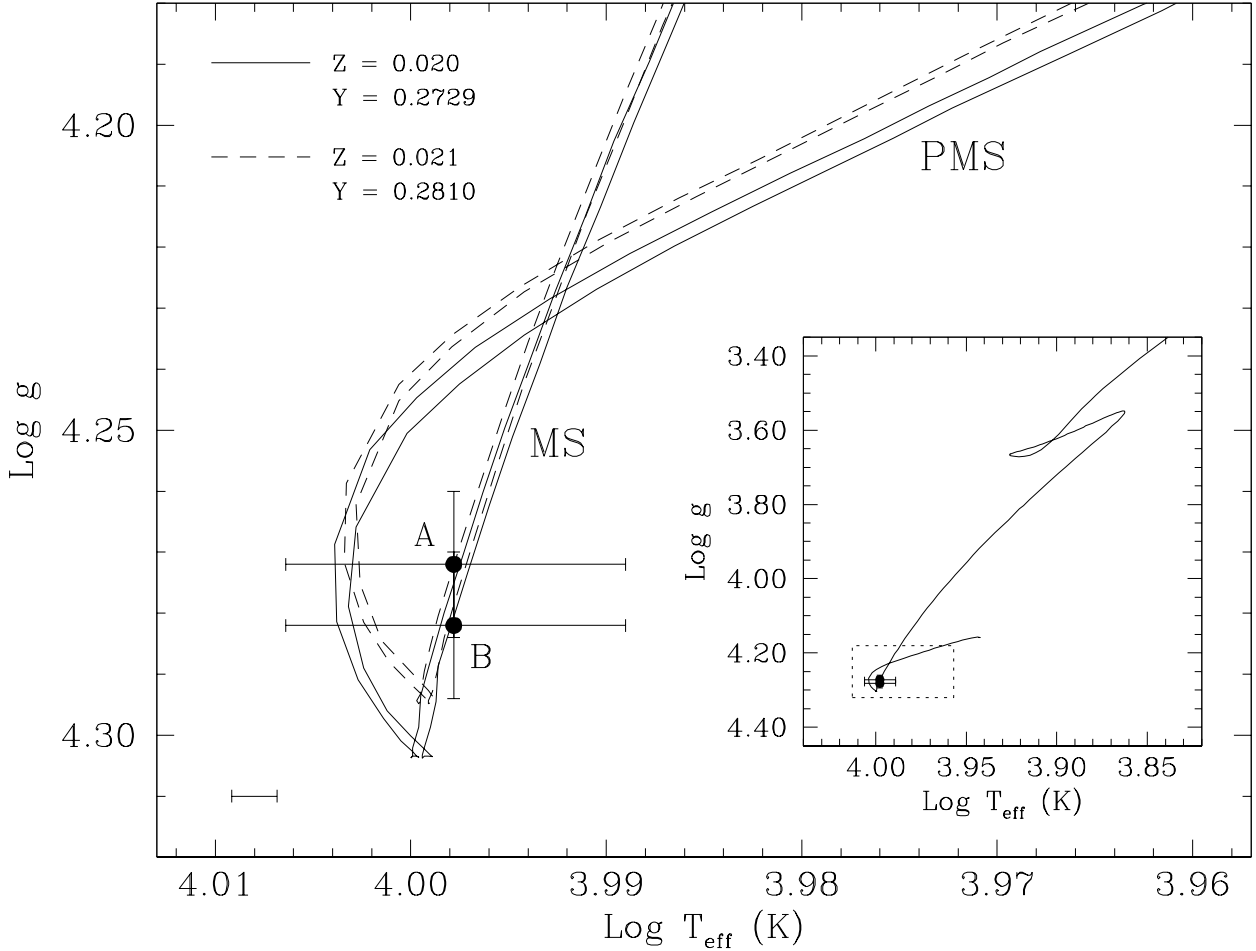}{5in}{270}{80}{80}{-320}{420}
\caption{Comparison with stellar evolution models
for both components in the $\log g$--$\log T_{\rm eff}$ plane.
Evolutionary tracks for two different compositions are shown, giving
good fits assuming GG~Ori is on the main sequence (MS). The
uncertainty in the placement of the tracks due to errors in the
measured masses is indicated with the error bar in the lower left.
The full extent of the main sequence band is shown in the inset,
emphasizing the fact that the system is very close to the
ZAMS.}
\end{figure}

\clearpage

\begin{figure}
\plotfiddle{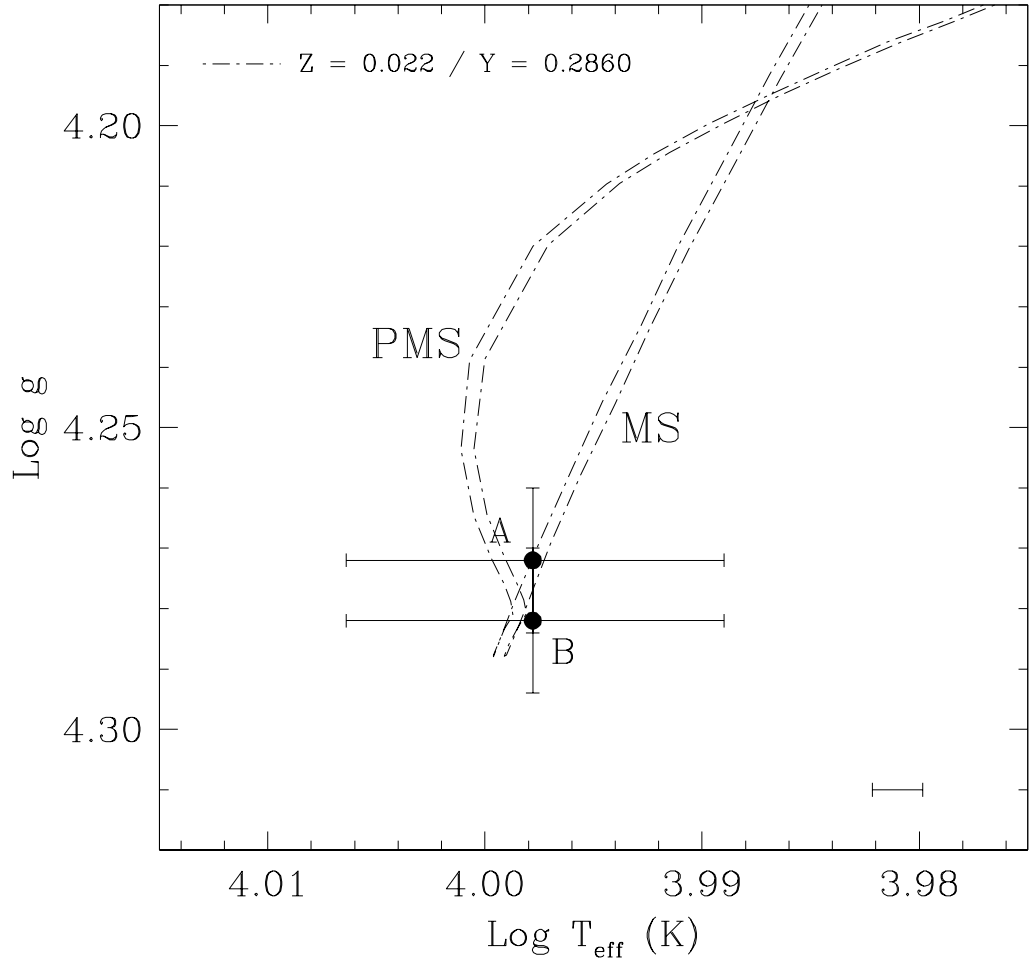}{4in}{0}{85}{85}{-270}{-160}
\caption{Best fit evolutionary tracks assuming
GG~Ori is still in the pre-main sequence (PMS) phase. The uncertainty
in the placement of the tracks due to errors in the measured masses is
indicated with the error bar in the lower right.}
\end{figure}

\clearpage

\begin{figure}
\plotfiddle{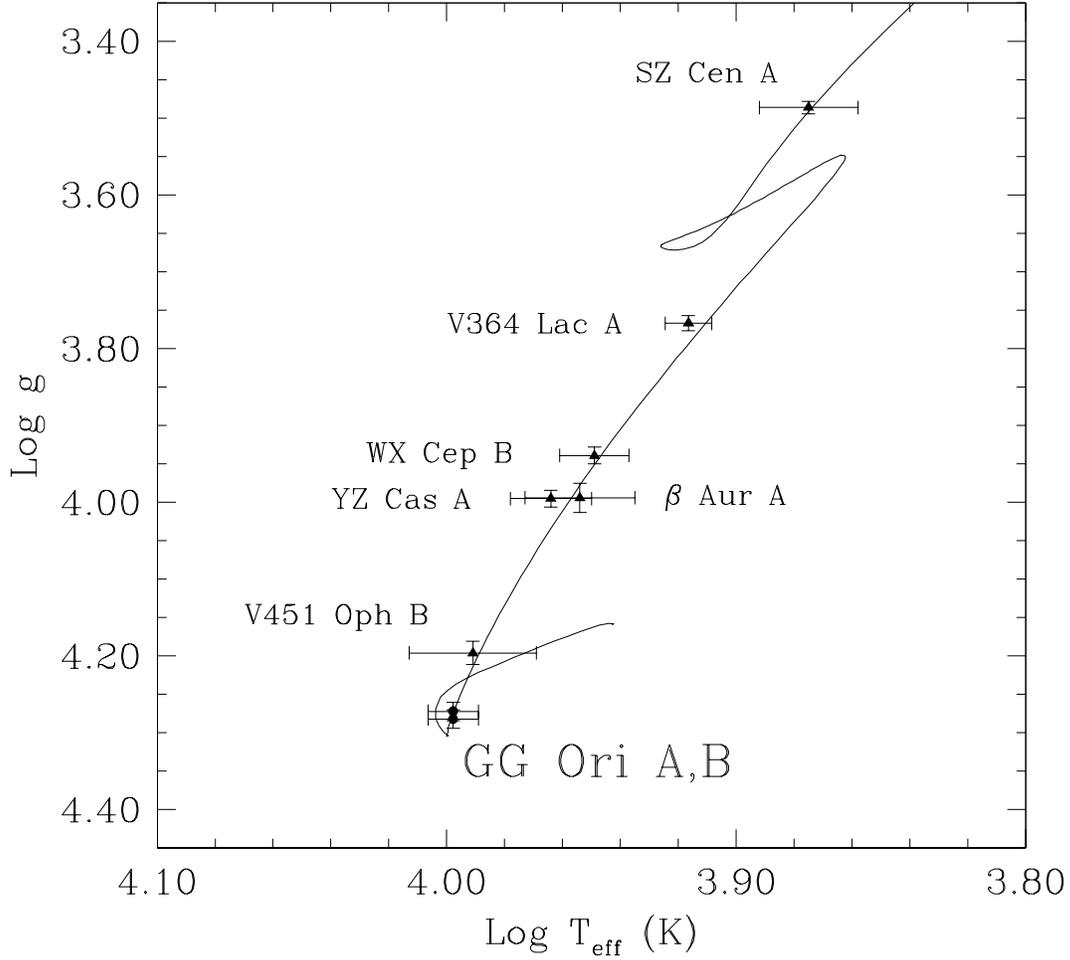}{4in}{0}{85}{85}{-270}{-160}
\caption{Evolutionary track for a mass equal to
that of the primary in GG~Ori and a composition close to solar ($Z =
0.020$, $Y = 0.2729$). Also shown are all other binary systems that
have accurate absolute dimensions available in which at least one
component has a similar mass. The 8 stars represented differ in mass
by less than 1\%, and all are adequately fit by this
model.}
\end{figure}

\clearpage
	
\begin{figure}
\plotfiddle{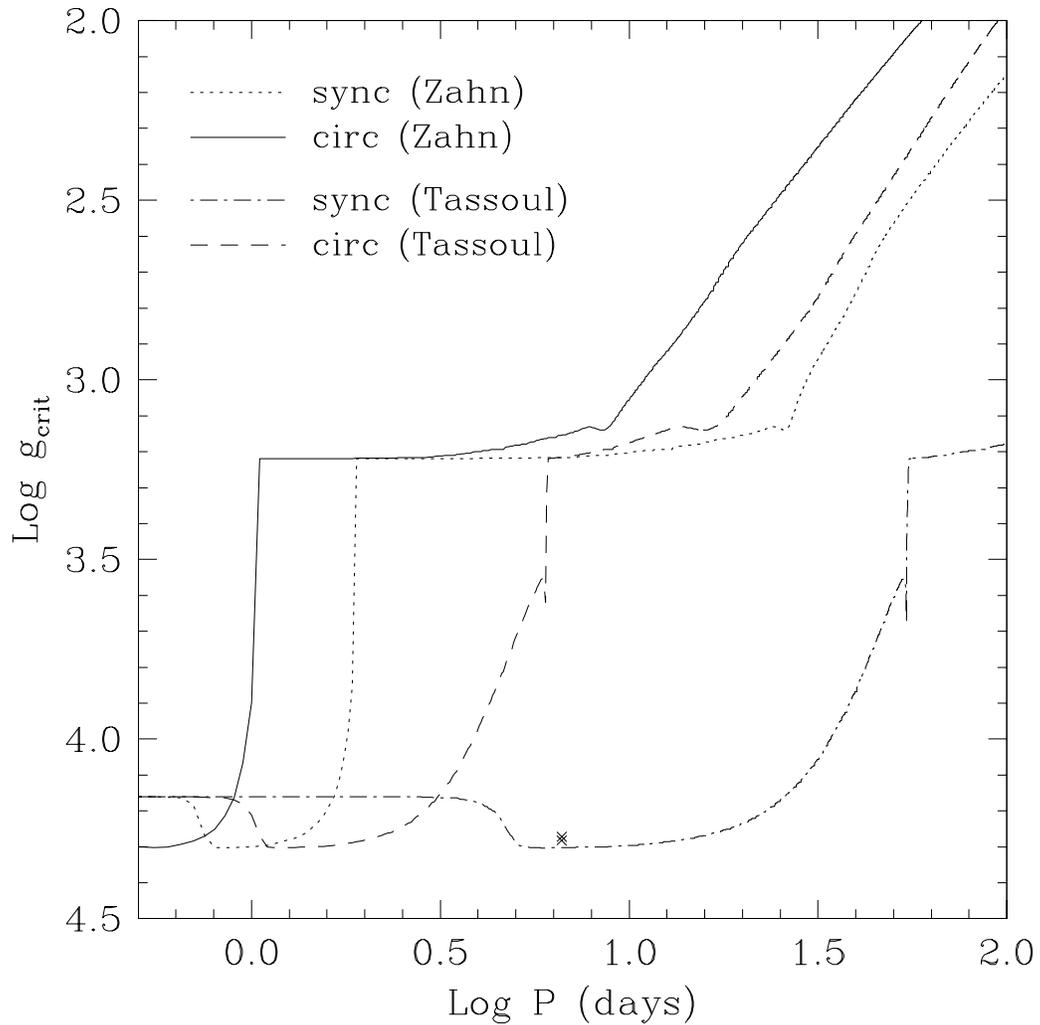}{6in}{0}{85}{85}{-270}{-160}
\caption{Critical values of the surface gravity at
which the stars in GG~Ori reach synchronization and circularization,
as a function of orbital period. Curves for two different tidal
mechanisms are shown: Zahn (1992) and Tassoul \& Tassoul (1997).
Evolution carries the stars upwards. Our measurements of $\log g$ for
the system are indicated with crosses. Both theories predict
circularization has not yet occurred (as observed), but they disagree
as to the synchronization.}
\end{figure}

\clearpage

\begin{figure}
\plotfiddle{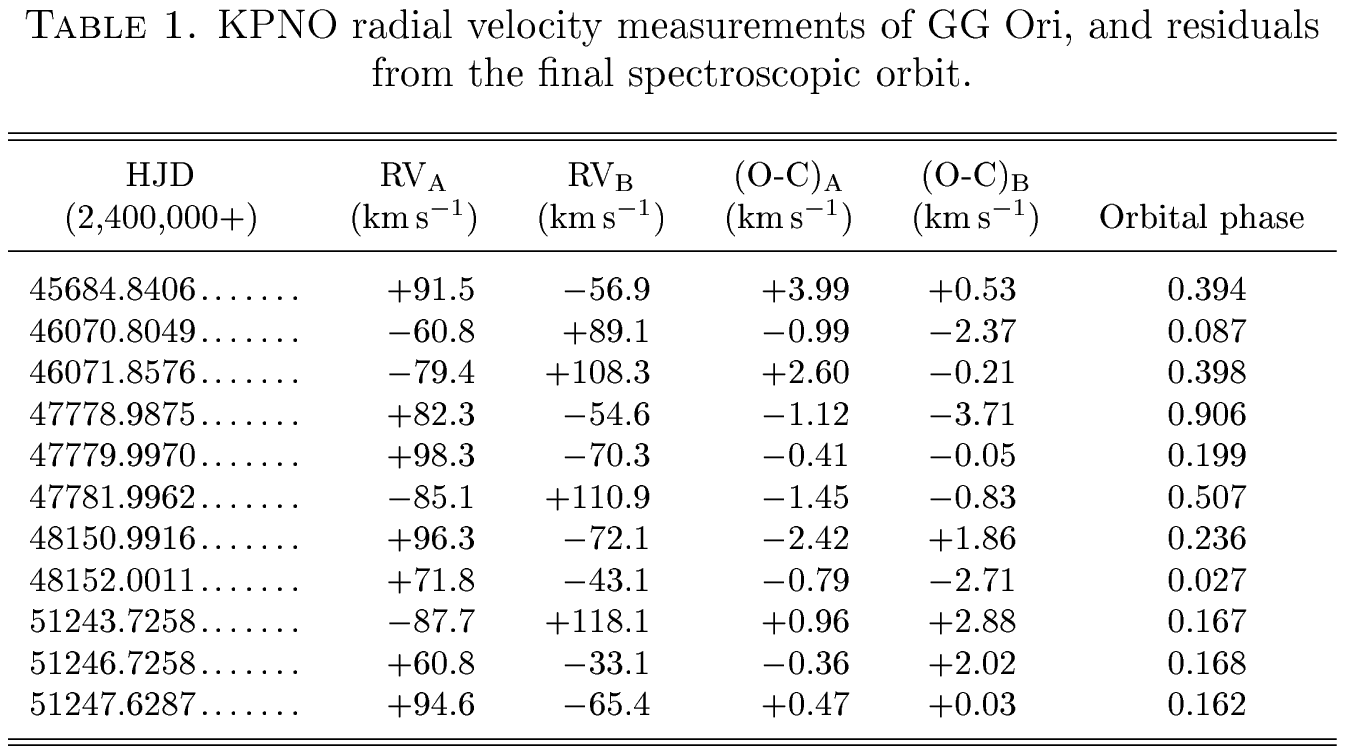}{8in}{0}{100}{100}{-300}{-40}
\end{figure}

\clearpage

\begin{figure}
\plotfiddle{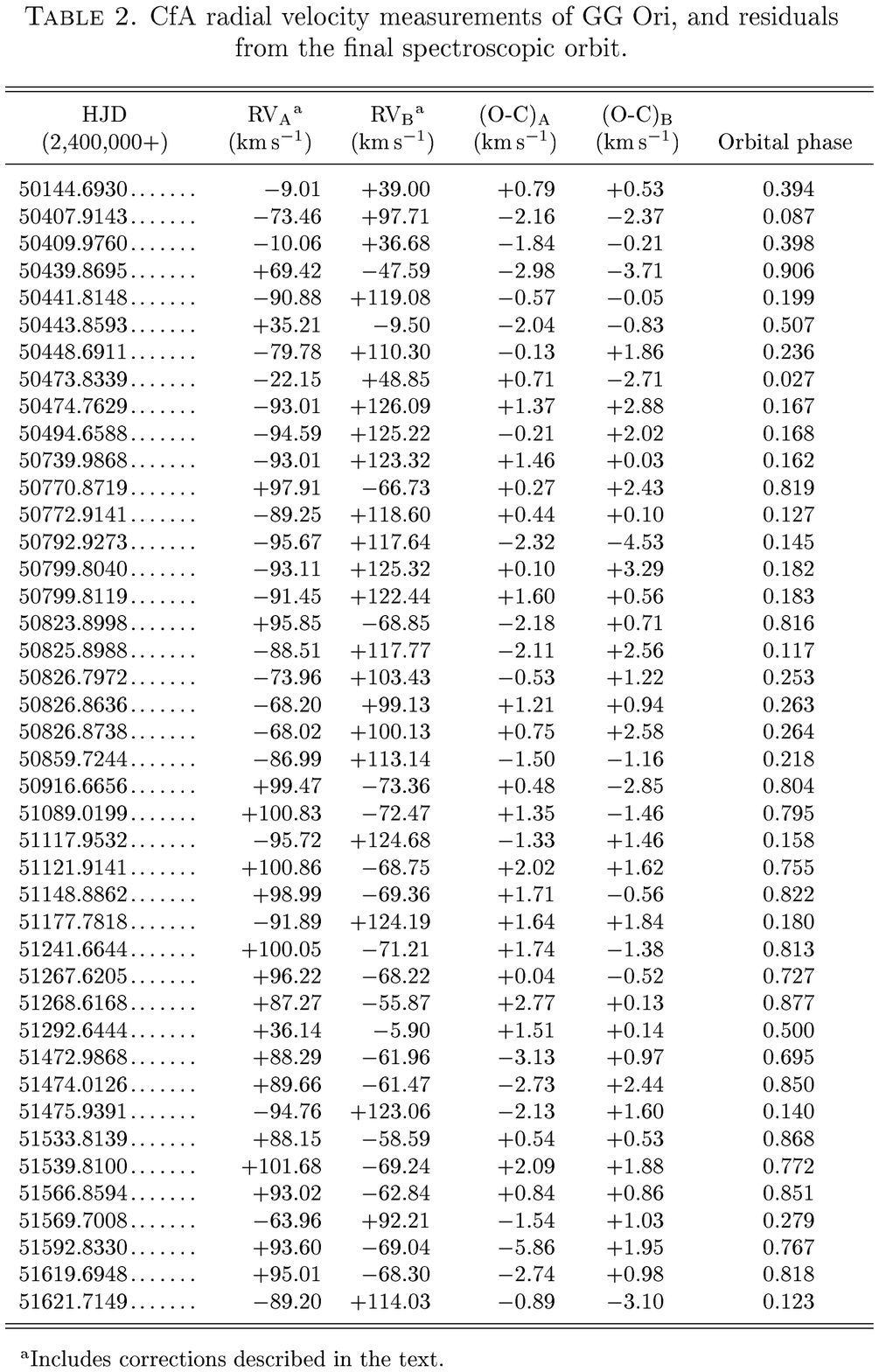}{11in}{0}{100}{100}{-300}{100}
\end{figure}

\clearpage

\begin{figure}
\plotfiddle{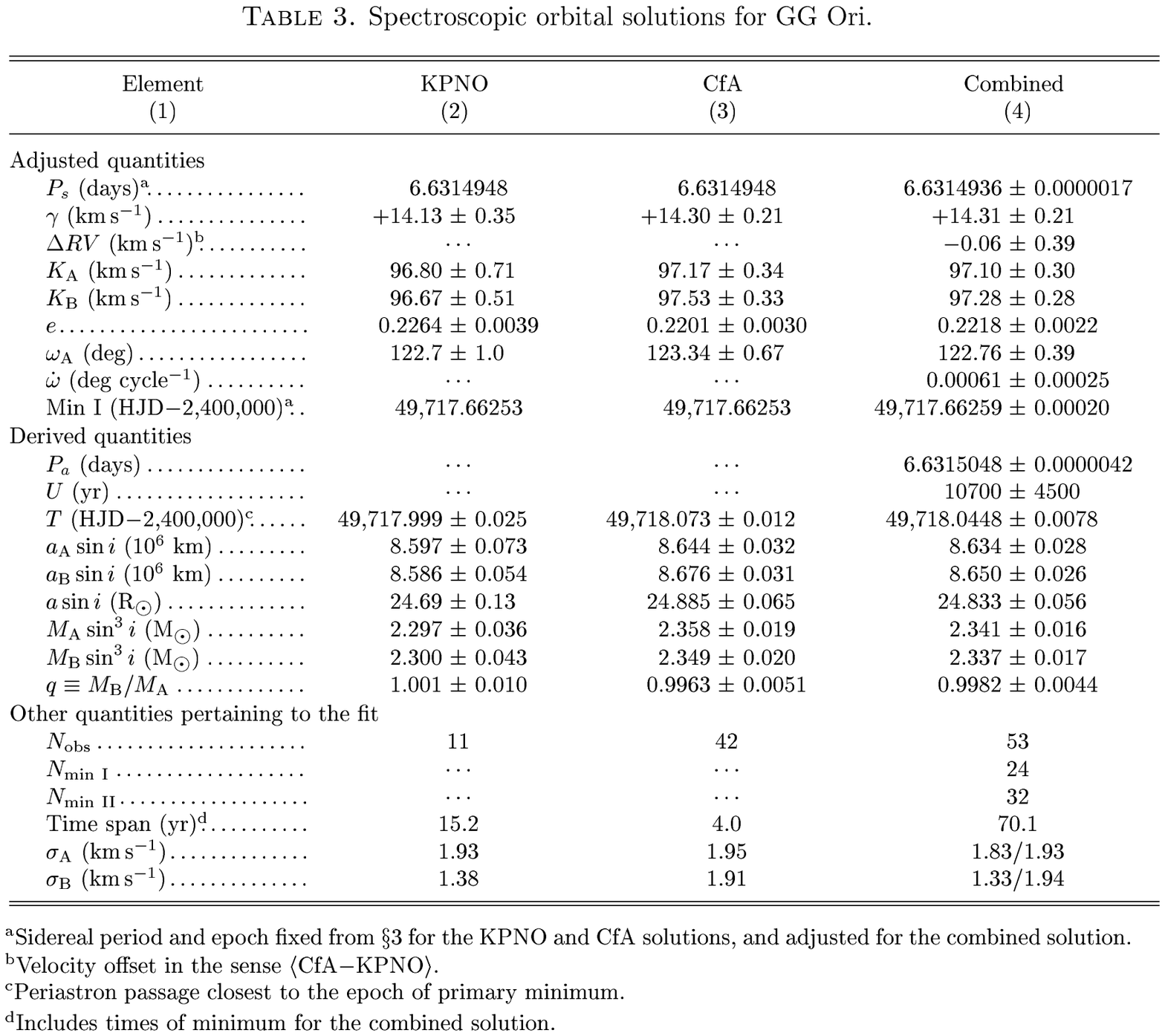}{8in}{0}{100}{100}{-305}{-40}
\end{figure}

\clearpage

\begin{figure}
\plotfiddle{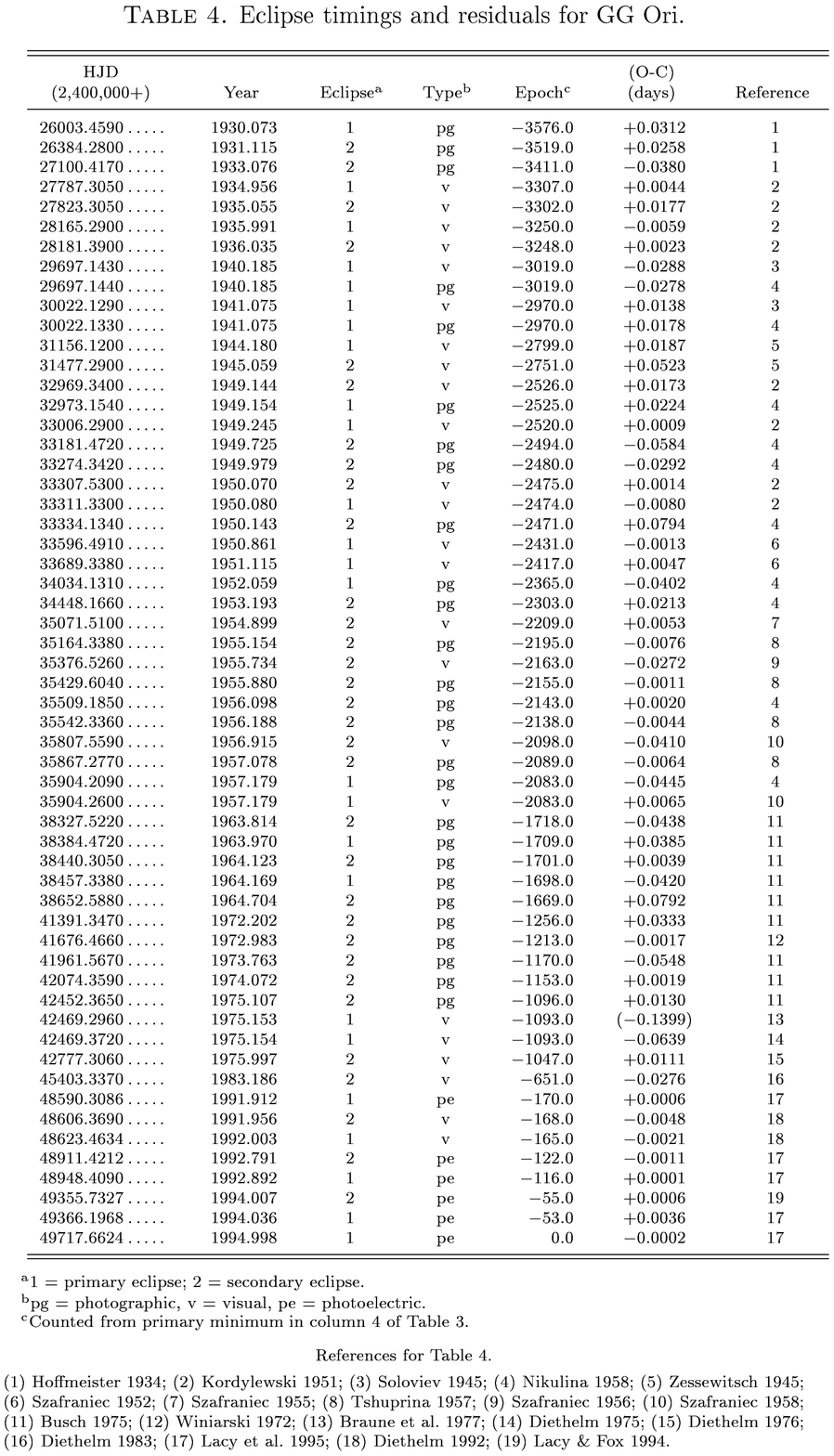}{11in}{0}{90}{90}{-275}{175}
\end{figure}

\clearpage

\begin{figure}
\plotfiddle{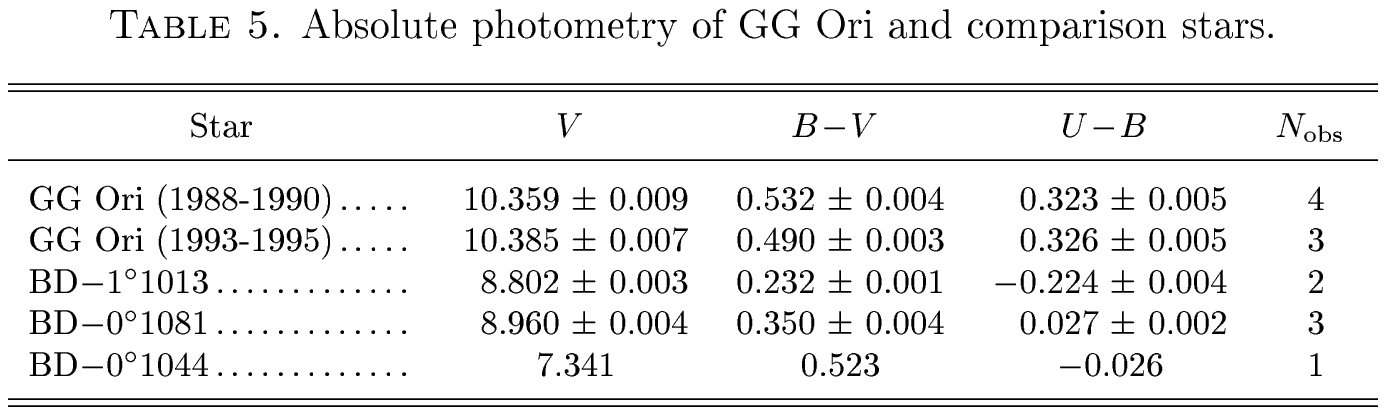}{8in}{0}{100}{100}{-300}{40}
\end{figure}

\clearpage

\begin{figure}
\plotfiddle{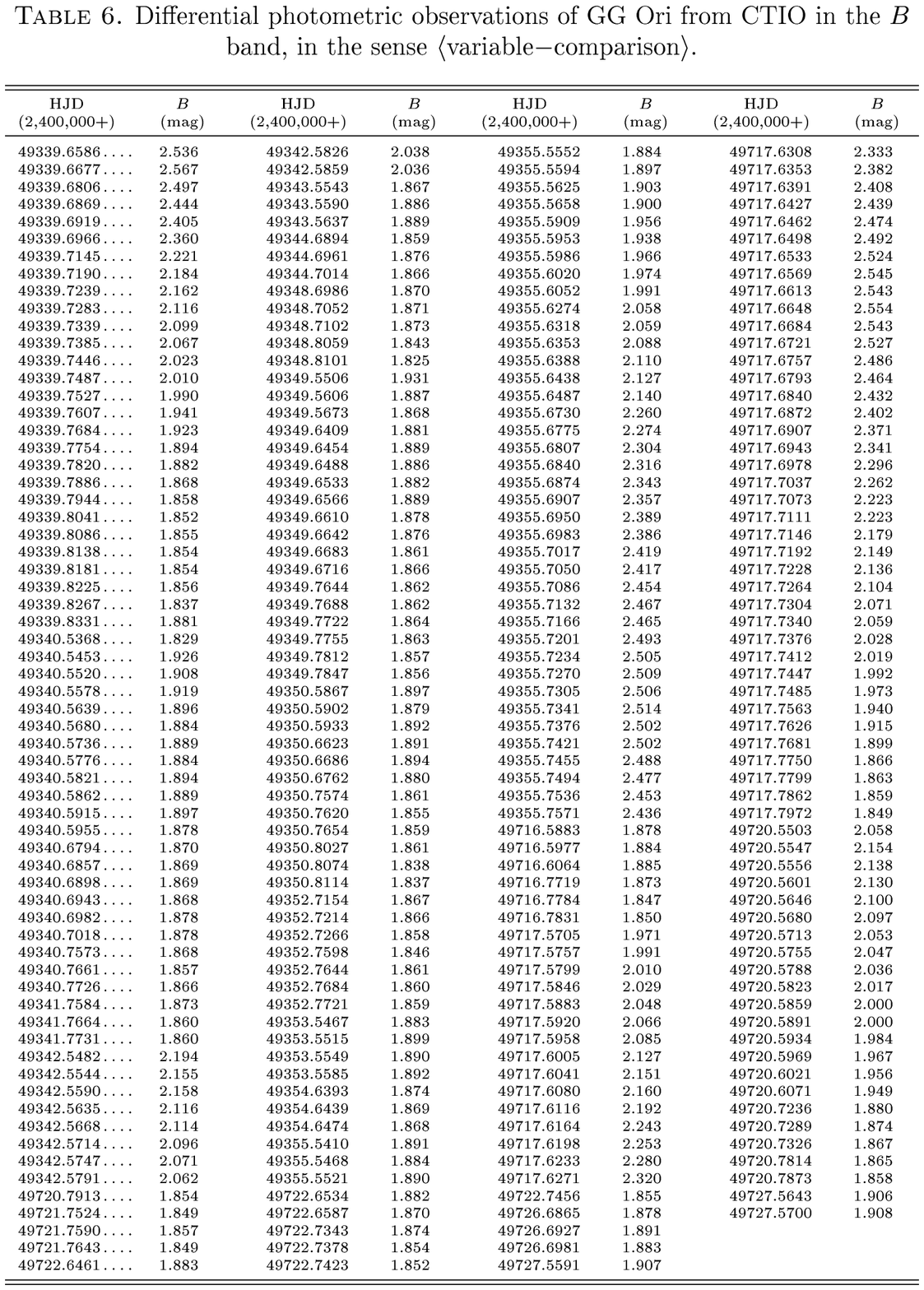}{11in}{0}{100}{100}{-290}{100}
\end{figure}

\clearpage

\begin{figure}
\plotfiddle{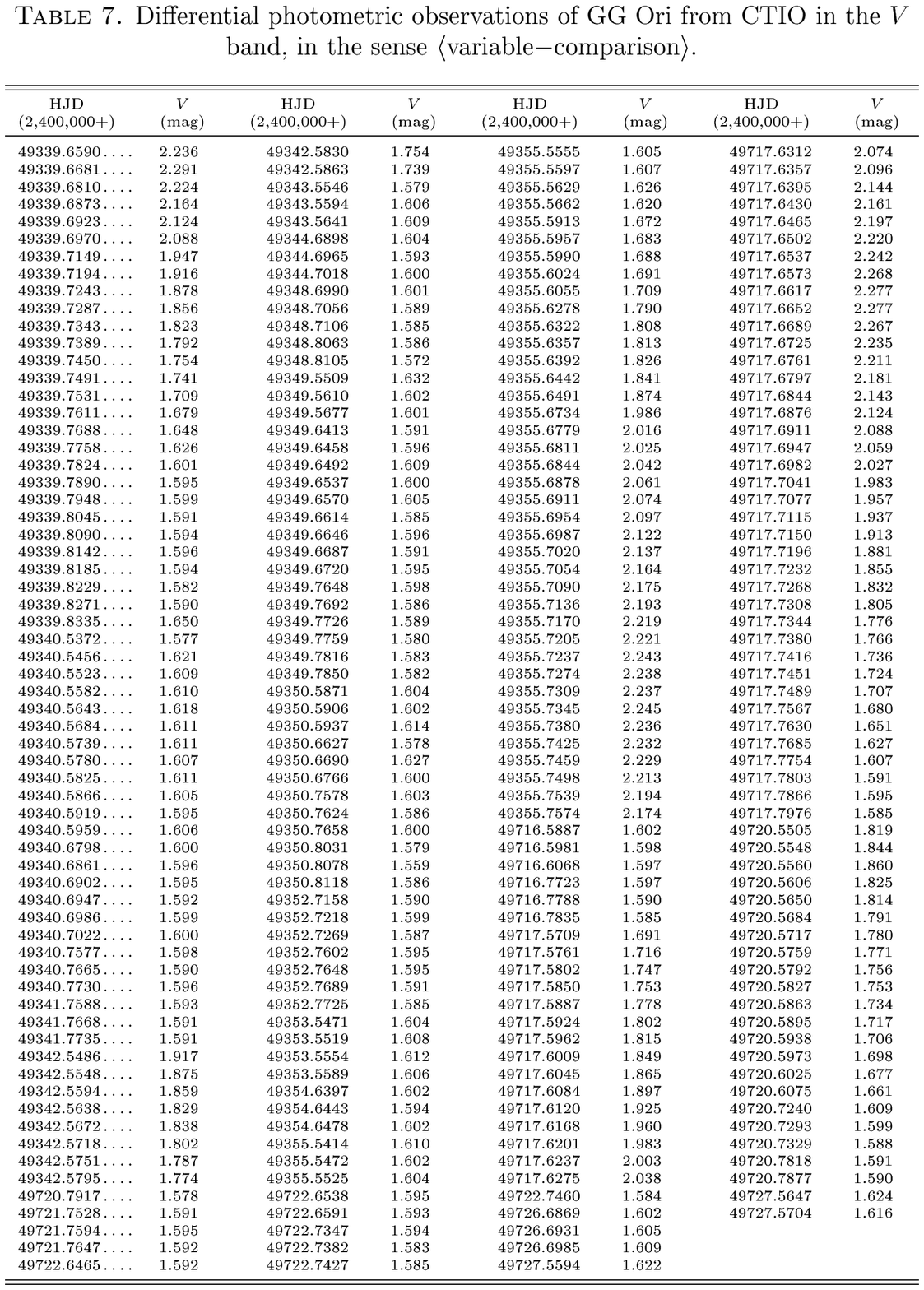}{11in}{0}{100}{100}{-290}{100}
\end{figure}

\clearpage

\begin{figure}
\plotfiddle{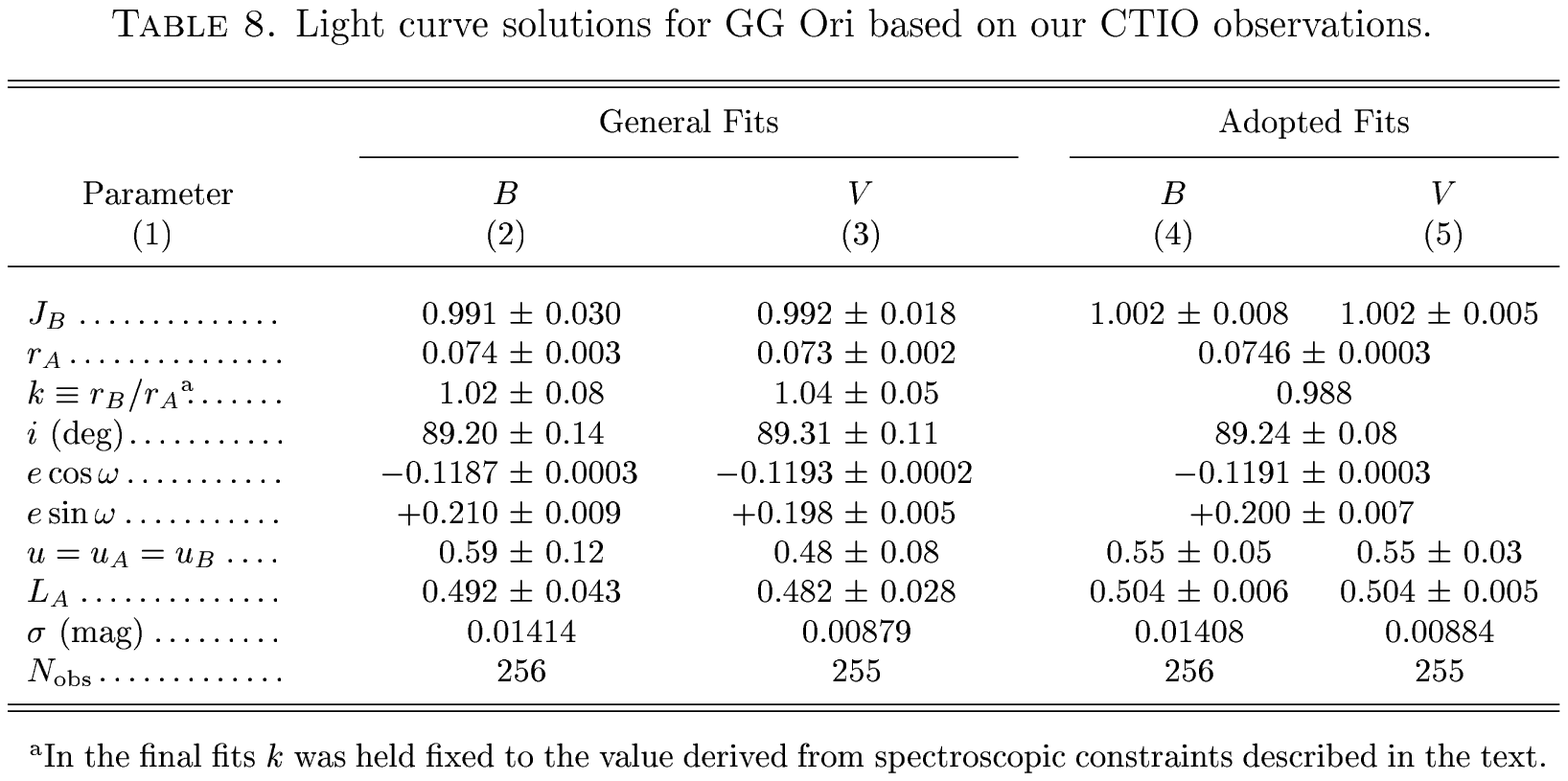}{8in}{0}{105}{105}{-320}{-50}
\end{figure}

\clearpage

\begin{figure}
\plotfiddle{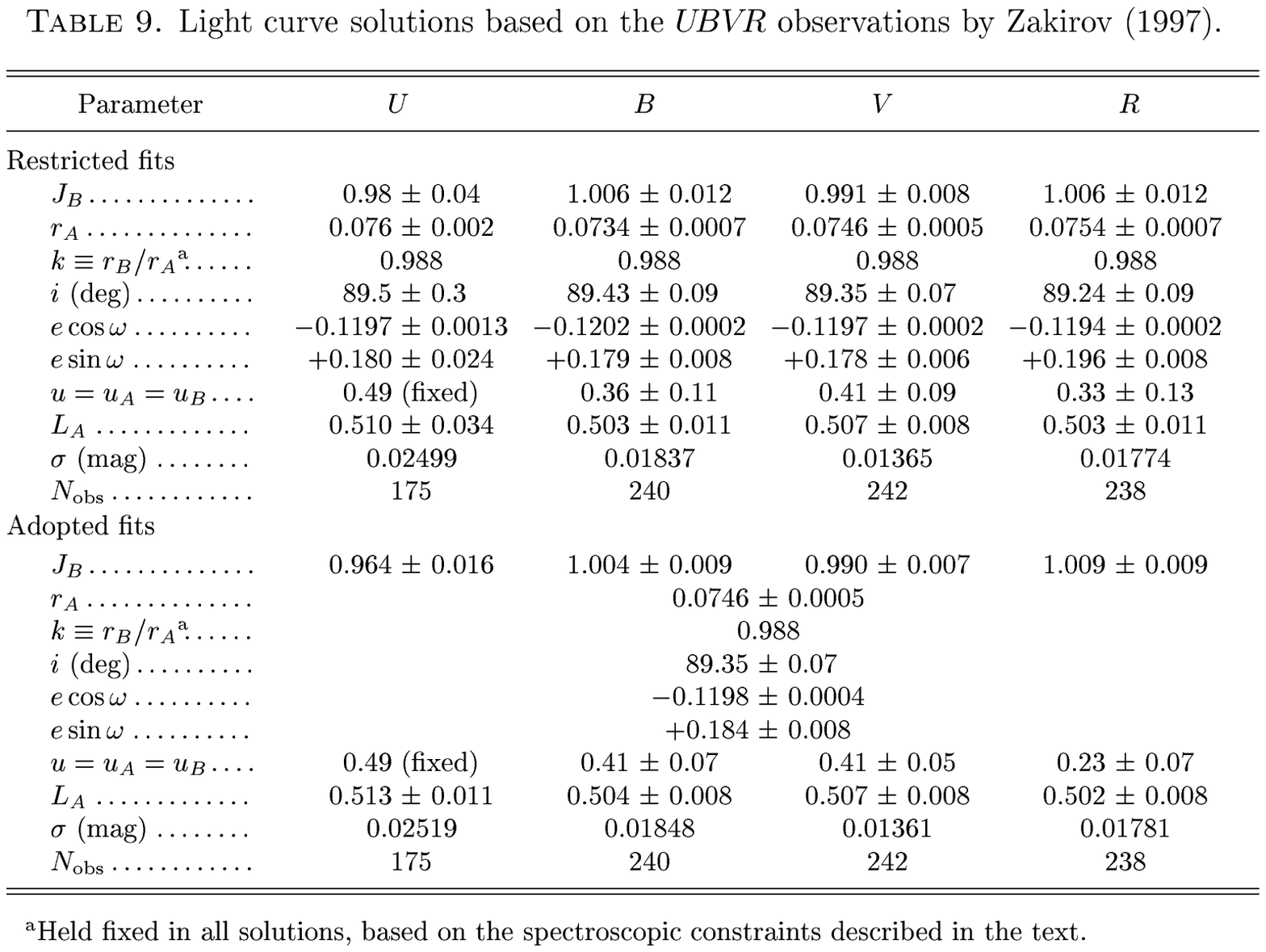}{8in}{0}{105}{105}{-315}{-100}
\end{figure}

\clearpage

\begin{figure}
\plotfiddle{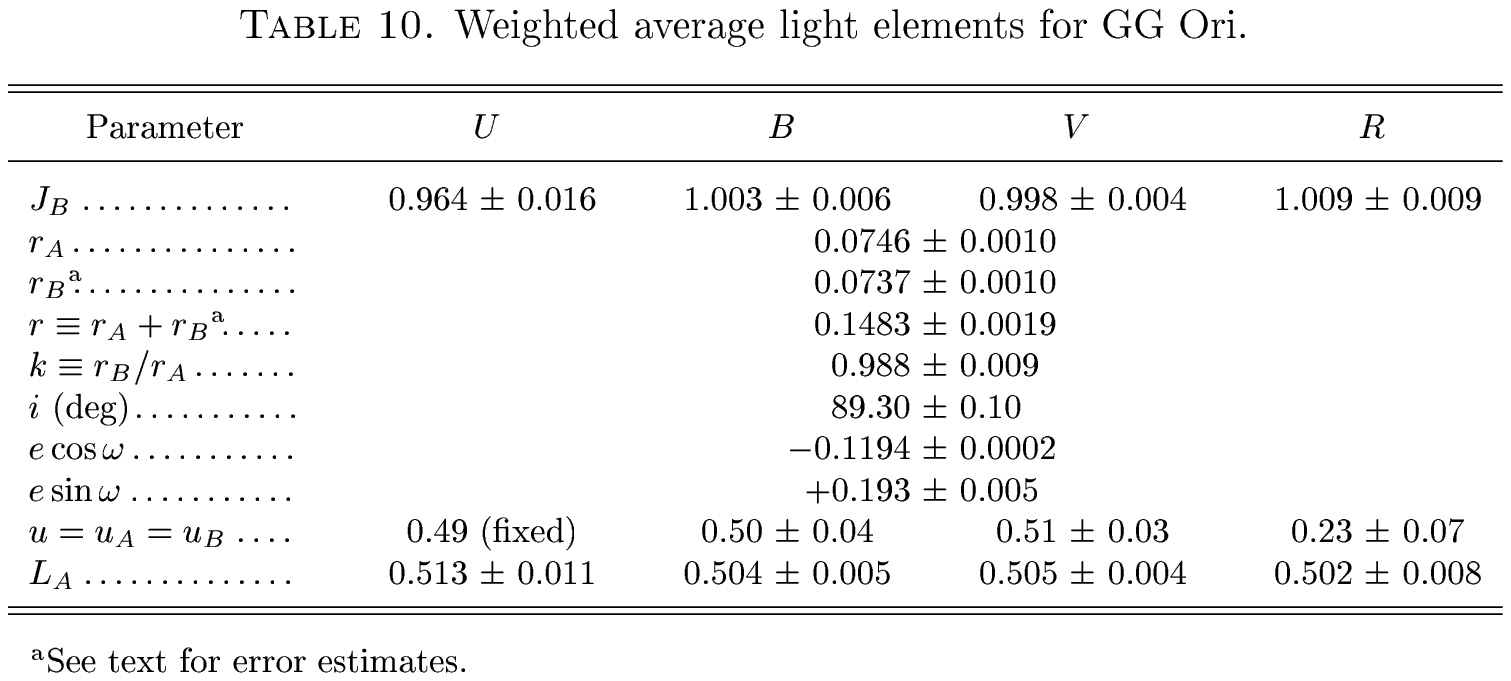}{8in}{0}{105}{105}{-320}{-100}
\end{figure}

\clearpage

\begin{figure}
\plotfiddle{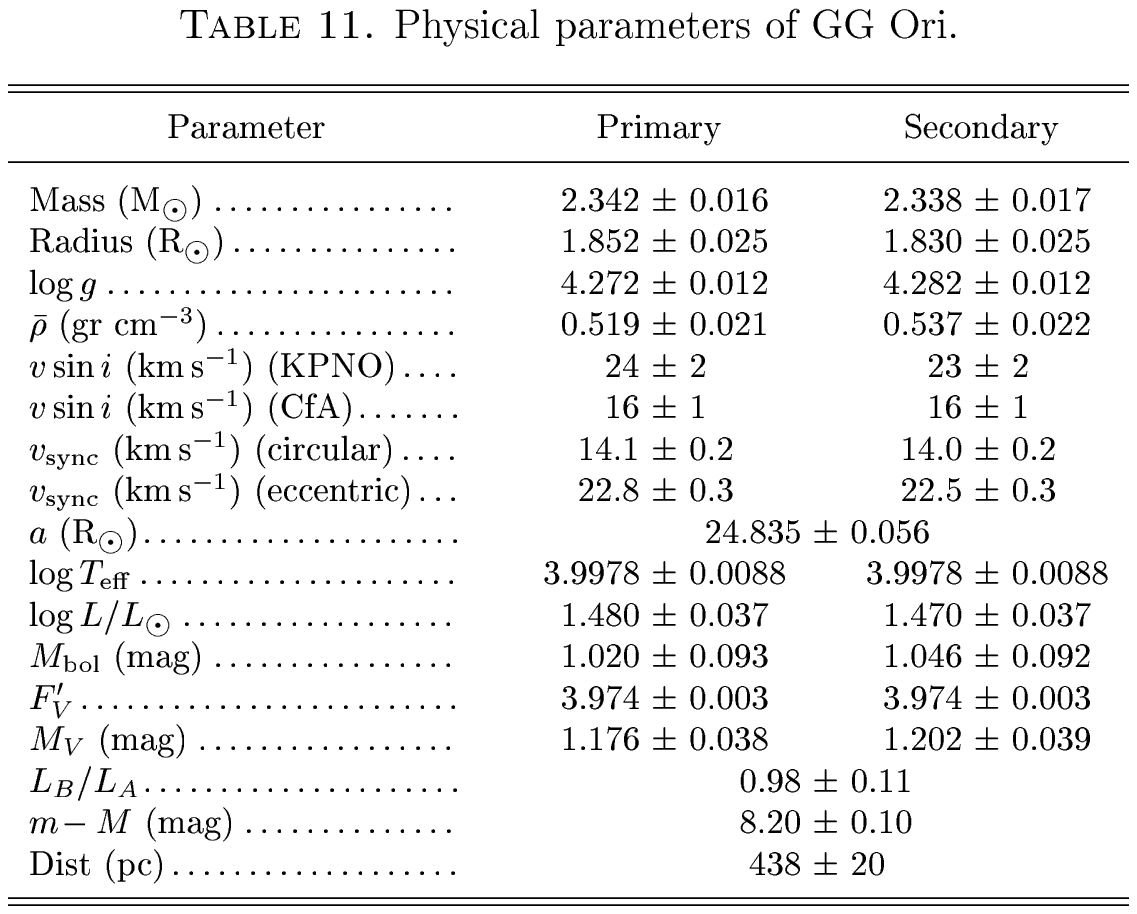}{8in}{0}{105}{105}{-320}{-100}
\end{figure}

\clearpage

\begin{figure}
\plotfiddle{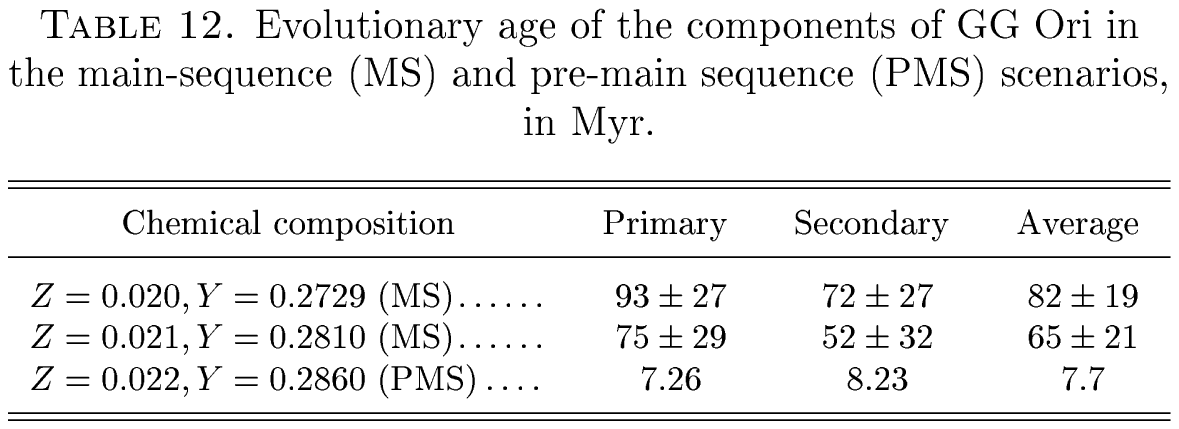}{8in}{0}{105}{105}{-320}{-100}
\end{figure}

\end{document}